\documentclass[12pt]{article}
\usepackage{epsfig,rotating,subeqn}
\begin{document}

\newcommand{\lsim}   {\mathrel{\mathop{\kern 0pt \rlap
  {\raise.2ex\hbox{$<$}}}
  \lower.9ex\hbox{\kern-.190em $\sim$}}}
\newcommand{\gsim}   {\mathrel{\mathop{\kern 0pt \rlap
  {\raise.2ex\hbox{$>$}}}
\lower.9ex\hbox{\kern-.190em $\sim$}}}
\def\be{\begin{equation}}
\def\ee{\end{equation}}
\def\ba{\begin{eqnarray}}
\def\ea{\end{eqnarray}}
\def\eps{{\varepsilon}}
\def\Ejet{E_{\rm jet}}
\def\tmin{t_{\rm min}}
\def\zmin{z_{\rm min}}
\def\Qmin{Q_{\rm min}}
\def\tsusy{t_{\rm SUSY}}
\def\Msusy{M_{\rm SUSY}}
\def\dlim{D_{\rm lim}}
\def\dmc{D_{\rm MC}}
\def\ap{\approx}
\def\bb{\leftarrow}


\title{\hfill \hfill {\small MPI-PHT-2003-28}\\
       \vskip1.0cm
{\bf Fragmentation functions in SUSY QCD and\\
     UHECR spectra produced in top-down models}}
     
\author{R. Aloisio$^{1}$, V. Berezinsky$^{1}$ and M. Kachelrie{\ss}$^{2}$\\
        {\it\small $^1$INFN, Laboratori Nazionali del Gran Sasso,
                   I--67010 Assergi (AQ), Italy} \\
        {\it\small $^2$Max-Planck-Institut f\"ur Physik 
                  (Werner-Heisenberg-Institut), 
                   D--80805 M\"unchen} }

\date{\today}
\maketitle

\abstract{
We present results from two different methods for the calculation of
hadron spectra in QCD and SUSY QCD with large primary
energies $\sqrt{s}$ up to $10^{16}$ GeV. The two methods considered
are a Monte Carlo (MC) simulation and the evolution of fragmentation 
functions described by the 
Dokshitzer-Gribov-Lipatov-Altarelli-Parisi (DGLAP) equations.
We find that the pion, nucleon and all-hadron spectra
calculated with the two methods agree well.
The MC simulation is performed with new hadronization functions
(in comparison with our previous work), motivated by 
low-energy ($\sqrt s < M_Z$) data 
and DGLAP. The hadron spectra calculated with both sets of
hadronization functions agree well, which indicates that our method
for calculating the hadronization function works successfully. The small
difference in the calculated hadron spectra characterizes the
uncertainties of this method.
We calculate also the spectra of photons, neutrinos and nucleons and compare
them with other published results. The agreement is good for all $x$ 
from $\sim 10^{-5}$ up to $x\leq 0.3$. The consistency of the
spectra calculated by different methods allows to consider the
spectral shape as a signature of models with decays or annihilations of
superheavy particles, such as topological defects or superheavy DM.
The UHECR spectra from these sources are calculated.}

\vskip0.8cm



\newpage

\section{Introduction}
\label{introduction}
Ultra-High Energy Cosmic Rays (UHECR) remain a puzzle in physics.
First of all, eleven AGASA \cite{AGASAsp} events with
$E\geq 1\times 10^{20}$~eV contradict the Greisen-Zatsepin-Kuzmin
(GZK) cutoff~\cite{GZK}, although the HiRes data~\cite{HiRes} are
generally considered to be consistent with the GZK cutoff~\cite{bw}.
If the UHECR primaries are protons (see below) and if they propagate
rectilinearly, as the claimed correlations with BL Lacs at lower
energies $(4 - 8)\times 10^{19}$~eV~\cite{TT} imply, then
their sources must be seen in the direction of the highest energy
events with $E \sim (2 - 3)\times 10^{20}$ eV detected by
HiRes~\cite{HiRes}, Fly's Eye~\cite{FE} and AGASA~\cite{AGASAhe}.
Indeed, the  proton attenuation length at
these  energies is only 20 -- 30 Mpc \cite{BGG1}, and the sources
should have been seen in the direction of these particles, since such
correlations
exist at considerably lower energies. This implies that particles with
$E\sim 10^{20}$ eV may have a different origin as those
with lower energies.

Meanwhile, there is strong evidence that primary particles at lower
energies, $1\times 10^{18}\lsim E \lsim (7-8)\times 10^{19}$~eV, 
are extragalactic protons, most probably from Active Galactic Nuclei
(AGN). The different pieces of evidence include: ({\it i\/}) Extensive
Air Shower (EAS) data confirm  protons as primaries~\cite{Sokol,Yak}, 
({\it ii\/}) the dip~\cite{dip}, seen with $\chi^2_{\rm dof}=0.7$   
in the spectra of AGASA, HiRes, 
Fly's Eye and Yakutsk~\cite{BGG2}, is a signature of the propagation of UHE 
protons in the extragalactic space and ({\it iii\/}) the beginning of
the GZK cutoff seen in the spectra of AGASA and HiRes~\cite{BGG2}.

According to the correlations with BL Lacs found in Ref.~\cite{TT} and
the analysis of small-angle clustering~\cite{clust}, protons should
propagate rectilinearly from the sources (AGN). However, if one
excludes the correlations 
with BL Lacs from the analysis, the propagation of protons in very
strong magnetic fields becomes also feasible~\cite{Sigl,jap}. 
Nevertheless, also in this case, the lack of a nearby source in the
direction of the highest energy events (e.g. at $E\sim 3\times
10^{20}$~eV) remains a problem for reasonable field strengths $B\sim 1$~nG:
the deflection angle, $\theta \sim l_{\rm att}/r_H=
3.7^{\circ} B_{\rm nG}$ given by the attenuation length $l_{\rm att}$
and the Larmor radius $r_H$, is small and sources should
be seen.

Many ideas have been put forward aiming to explain the observed
superGZK ($E \gsim (6-8)\times 10^{19}$~eV) events: strongly interacting 
neutrinos~\cite{nu} and new light hadrons~\cite{gluino} as unabsorbed
signal carriers, $Z$-bursts~\cite{Z}, 
Lorentz-invariance violation \cite{Lorentz}, Topological Defects (TD)
(see \cite{TD} for a review), and Superheavy Dark Matter (SHDM)
(see \cite{SHDM} for a review).

The two last models listed above share a common feature: UHE particles are
produced in the decay of superheavy (SH) particles or in their
annihilation. In the case of TD they are unstable 
and in the case of SHDM  long-lived particles. 
We shall call them collectively $X$ particles. In the $Z$-burst model,
the decay of a much lighter particle, the $Z$-boson, is involved, while
the decay products are boosted by very large Lorentz
factors. Annihilation takes place in the case
of monopolonia \cite{monop}, necklaces \cite{neckl} and SHDM particles
within some special models \cite{BDK}. As elementary particle physics
is concerned, both processes proceed in a way similar to
$e^+e^-$~annihilation into hadrons: two or more off-mass-shell
quarks and gluons are produced and they initiate QCD cascades. 
Finally the partons are hadronized at the confinement radius. 
Most of the hadrons in the final state are pions and thus the typical
prediction of all these models is the dominance of photons at the
highest energies $E \gsim (6-8)\times 10^{19}$~eV. 

This prediction is questioned by the AGASA data~\cite{AGASA-gamma} at
$E \gsim 1\times 10^{20}$~eV combined with the
recent recalculation \cite{AhPl} of the muon number in
photon-initiated EAS at the highest
energies. Previously~\cite{AhKa,Capd}, the muon content in
photon-induced showers was found to be similar to that in
proton-induced showers. According to the new calculations, the muon
number is still large, but  5 -- 10 times smaller than
in hadronic showers. From eleven AGASA events at $E\gsim 1\times 10^{20}$~eV, 
the muon content is measured in six  and in two of them it is
high. The muon content of the remaining four events is close to the one
in photon-induced showers. MC simulations \cite{AGASA-gamma} show that
a primary flux comprised {\em completely\/} by photons is
disfavored. We shall discuss further these data in the Conclusions. 

The spectrum of hadrons produced in the decay/annihilation of
$X$ particles is another signature of models with superheavy
$X$ particles.  Several methods for the calculation of these spectra
have been developed in the past several years.

The mass of the decaying particle, $M_X$, or the energy of annihilation
$\sqrt{s}$, is in the range $10^{13}$ -- $10^{16}$~GeV. 
The existing QCD MC codes become numerically unstable at much smaller
energies, e.g., at $M_X \sim 10^7$~GeV in the case of HERWIG. Moreover,
the computing time increases rapidly going to larger energies.
Nevertheless, one of the first spectrum calculation has been
performed with the help of HERWIG for energies up to $M_X \sim 10^{11}$~GeV,
and the computed spectra were extrapolated to
$M_X \sim 10^{13}$~GeV~\cite{Sar}. Supersymmetry was not included in
these calculations.

Another option used in the first calculations is the {\em limiting
spectrum\/}, an analytic method developed in Refs.~\cite{lim}. 
This method has been found to be  very successful at LEP energies
(see \cite{Khoze} and references therein). Two basic assumptions are
involved in this method:  ({\it i\/}) the beta function $\beta$ 
describing the 
running of the QCD coupling $\alpha_s$  is taken to be constant,
i.e. $\alpha_s(k_{\perp}^2) \propto 1/\ln(k_{\perp}^2/\Lambda^2)$ for
all transverse momenta $k_{\perp}$, and ({\it ii\/}) the minimum
virtuality $Q_0$ of partons, down to which the cascade develops
perturbatively, is taken equal to the scale $\Lambda$. The high energy
supersymmetric generalization of this solution has been obtained in
Ref. \cite{BKlim}. Later, a comparison with the MC simulation~\cite{BKMC}
showed that the limiting spectrum does not describe well hadron
spectra in SUSY QCD and that the assumption ({\it i\/}) is mainly 
responsible for the discrepancies found. Indeed, changing the
evolution of $\alpha_s(k_{\perp}^2)$ an agreement of the spectra
around the Gaussian peaks was obtained~\cite{BKMC} when in the SUSY
QCD MC $\alpha_s$ with $\beta=$~const. was used and $\alpha_s$ was 
frozen at small $k_{\perp}^2$, which is a reasonable physical 
assumption, e.g., in the DGLAP method%
\footnote{We are not able to find any place in our paper
\cite{BKMC} where according to \cite{SaTo} we ``acknowledge errors.''}.
For more details see Ref.~\cite{BKMC}. 

Monte Carlo simulations are the most physical approach for high
energy calculations which allow to incorporate many important physical
features as the presence of SUSY partons in the cascade and
coherent branching~\cite{MW}. The perturbative part of our MC
simulation is similar to other existing MC programs and hence reliable. 

For the non-perturbative
hadronization part a phenomenological approach is used in
Ref.~\cite{BKMC}. The fragmentation of parton $i$ into a hadron $h$ is
expressed through perturbative fragmentation function of partons
$D_i^j(x,M_X)$ convoluted  with hadronization functions $f_j^h(x,Q_0)$
at the scale $Q_0$,
\be
D_i^h(x,M_X)=\sum_{j=q,g}\int_x^1\frac{dz}{z}D_i^j(x/z,M_X,Q_0)f_j^h(z,Q_0)\,,
\label{hfunc}
\ee
where the hadronization functions do not depend on the scale
$M_X$ (as notation for the scale in this paper we shall use also
$\sqrt s=M_X$ and $t=\ln(s/s_0)$).
This important property of hadronization functions allows us
to calculate $f_i^h(x,Q_0)$ from available LEP data, $D_i^h(x,M_X)$
at the scale $M_X=M_Z$, and then to use it for the calculation of
fragmentation functions $D_i^h(x,M_X)$ at any arbitrary scale $M_X$. 
Our approach reduces the computing time compared to usual MC
simulations and allows the fast calculation of hadron spectra 
for large $M_X$ up to $M_{\rm GUT}$. The MC of Ref.~\cite{BKMC}
has two versions: one for ordinary QCD and another one for SUSY QCD.

The perturbative part of the MC simulation in Ref.~\cite{BKMC} includes
standard features such as  angular ordering, which provides the
coherent branching and the correct Sudakov form factors, as well as SUSY
partons. Taking into account SUSY partons results only in small
corrections to the production of hadrons, and therefore a simplified
spectrum of SUSY masses works with good accuracy.

The weak influence of
supersymmetry is explained by the decay of SUSY partons, when the
scale of the perturbative cascade reaches the SUSY scale
$Q_{\rm SUSY}^2 \sim 1~{\rm TeV}^2$. Most of the energy of SUSY partons
remains in the cascade in the form of energy of ordinary partons, left
after the decay of SUSY partons.  The qualitatively new
effect caused by supersymmetry is the effective  production of the
Lightest Supersymmetric Particles (LSP), which could be neutralinos or
gluinos. The spectra of neutralinos were calculated in \cite{BKMC}.

MC simulations allow also to calculate the characteristic feature of
the QCD spectrum, the Gaussian peak, which is beyond the power of the
next method we shall review.

The fragmentation functions $D_i^h(x,M_X)$  at a high scale $M_X$  can
be calculated 
evolving them from a low scale, e.g. $M_X=M_Z$, where they are known
experimentally. This evolution is described by the
Dokshitzer-Gribov-Lipatov-Altarelli-Parisi (DGLAP) equation \cite{GLD,AP}
which can be written schematically as
\be
\partial_t D_i^h=\sum_j\frac{\alpha_s(t)}{2\pi}P_{ij}(z)\otimes
D_j^h(x/z,t)\,,
\label{DGLAP}
\ee
where $t=\ln(s/s_0)$, $\otimes$ denotes the convolution
$f\otimes g=\int_z^1 dx/x f(x)g(x/z)$,
and $P_{ij}$ is the splitting function which describes the emission of
parton $j$ by parton $i$. 
Apart from the experimentally rather well
determined quark fragmentation function $D_q^h(x,M_Z)$, also the gluon
FF $D_g^h(x,M_Z)$ is needed for the evolution of
Eq.~(\ref{DGLAP}). The gluon FF can be taken either from MC
simulations or from fits to experimental data, in particular
to the longitudinal polarized $e^+e^-$ annihilation cross-section and
three-jet events.

The first application of this method for the calculation of hadron spectra
from decaying superheavy particles has been made in Refs.~\cite{Rub,FoKa}, 
followed by Refs.~\cite{SaTo,Co,CoFa,To,BaDr1,BaDr2}. The most detailed 
calculations have
been performed in Ref.~\cite{BaDr2}, where more than 30 different
particles were allowed to be cascading and the mass spectrum of the
SUSY particles was taken into account. Although at $M_Z$,  which is
normally the initial scale in the DGLAP method, the fragmentation
functions for supersymmetric partons are identically zero, they can be
calculated at larger scales $t$: 
SUSY partons are produced above their mass threshold, when their
splitting functions are included in Eq.~(\ref{DGLAP}).
In this work we prove that this method is correct.
Also, the LSP spectrum can be computed within the DGLAP approach 
\cite{BaDr2,IbTo}. 

A problem of the DGLAP method are the fragmentation functions at small
$x=2p/\sqrt{s}$, where $p$ is the momentum. 
The DGLAP equations allow to evolve fragmentation functions known at
$x>x_{\rm min}$ from a starting scale $\sqrt{s_0}$ to a higher scale
$\sqrt{s}$.  Therefore, $x_{\rm min}$ of the FF at the starting scale 
$\sqrt{s_0}$ determines the $x$ range accessible for {\em all\/} $s$. 
Requiring that perturbation theory can be used, $p_{\perp}>\Lambda
\sim 0.25$~GeV, leads for the starting scale $M_Z$ to $x> 0.005$. 

In Ref.~\cite{BaDr1,BaDr2} the initial
fragmentation functions are taken from Ref.~\cite{Kniehl} at the scale 
$Q_0 \sim 1.4$~GeV and are extrapolated to very low $x\sim 10^{-5}$,
i.e. into the non-perturbative region. The formal DGLAP
evolution between the scales $Q_0$ and $M_Z$ is described by equations 
with $\alpha_s(s)$ not depending on $x$ (see e.g. Eq.~(\ref{equ}) in
Section \ref{DGLAP_sec}). Surprisingly enough this method works well:
The fragmentation functions for $x$ as low as $10^{-5}$ evolved 
in Ref.~\cite{BaDr1,BaDr2} to large $M_X$ coincide with our MC
simulation.

Near the GUT scale all three gauge couplings $\alpha_i$ ($i=1,~2,~3$)
have approximately equal numerical values. Naively one expects that
at scales $M_X$ close enough to the GUT scale all particles
including e.g. leptons and electroweak (EW) gauge bosons are cascading like
QCD partons. 

In fact, the influence of the masses of the EW gauge bosons
on soft singularities has to be carefully studied. In the MC approach,
the leading effect of the finite masses of the gauge boson can be
implemented in a rather  
straightforward way~\cite{BKO}. Cascading of the longitudinal modes
of the EW gauge bosons or of the Higgs bosons is a subleading effect.
In Ref.~\cite{BKO}, it was demonstrated that EW cascading occurs at $M_X\gsim
10^{6}$~GeV, even if only leptons and EW bosons are included in the
consideration. The interactions with quarks and gluons mixes EW and QCD
cascades. If, for example, a hypothetical $X$ particle couples
at tree level only to neutrinos, their further cascading results in the
production of the QCD partons and thus of hadrons. Thus the production
of neutrinos through decays of $X$ particles with such couplings is
constrained by the usual electromagnetic (e-m) cascade limit in the
universe. EW cascading has been included in the calculations of
Refs.~\cite{BaDr1,BaDr2}, though in a formal way.

In this paper we shall study the agreement of two methods: MC
and DGLAP equations for the calculation of spectra produced in the decay
or annihilation of superheavy particles. For this we shall calculate
the spectra using the same assumptions in both methods. We shall also
compare the results obtained by different groups and confront the
calculated spectra with recent ones measured by UHECR experiments.

The paper is organized as follows: in Section 2 we introduce the DGLAP
equations and the technique used to solve them. In Section 3 we
discuss the properties of the hadronization functions used in the MC
simulation and obtain new low-energy motivated hadronization functions
as extension of those used in \cite{BKMC}. Then we compare the
fragmentation functions calculated with the MC and the DGLAP equations
in Section 4. Photon, neutrino and proton spectra, needed for UHECR
calculations, are computed in Section 5 and compared with the spectra
obtained in Refs.~\cite{SaTo,BaDr1,BaDr2}. 
Finally, in Section 6 we consider the consequences of our results 
for models of superheavy DM and Topological Defects.

\section{DGLAP equations in SUSY QCD}
\label{DGLAP_sec}
If the $X$ particle decays
into partons $i=u,\bar u,\ldots,g$, the parton FFs $D_i^h(x,m_X^2)$ can be
defined as the probability of fragmentation of a parton $i$ into a hadron $h$
with momentum fraction $x=2p/M_X$. The evolution of FFs with increasing scale 
$s=M_X^2$ (or $t=\ln s/s_0$)  is governed by the DGLAP equations,
\be
 \partial_t D_j^h(x,t)= \sum_j
 \int_x^1  \frac{dz}{z} D_i^h(x/z,t)P_{i\to j}(z,t) \,.
\label{DGLAP1}
\ee
Multiplying Eq. (\ref{DGLAP1}) by $x$ and integrating it over $x$ shows
that the DGLAP equations conserve momentum, 
$\partial_t\int D_j^h(x,t)xdx=0$, if $\sum_j\int P_{i\to j}(x)xdx=0$.

Since in the limit $s\gg m_q^2$ all
quark flavors couple to gluons in the same way, the gluon FF mixes
only with the flavor singlet FF of quarks,
\be
D_q^h(x,t) = \frac{1}{n_f}\sum_{\rm flavors}
             \left( D^h_{q_i}(x,t) + D^h_{\bar q_i}(x,t) \right) \,,
\ee
where the summation goes over the number of active quark flavors
$n_f$ involved in the process ($n_f$ increases with increasing 
scale $t$).

The coupled evolution equation for the gluon FF $D_g^h$ and the
quark singlet FF $D_q^h$ becomes then
\be \label{matrix1}
 \partial_{t}\left(\begin{array}{c}
  D_q^h(x,t)\\
  D_g^h(x,t)\\
                  \end{array} \right)=
\left[
\left( \begin{array}{cc}
       P_{qq}(x,t) &  P_{gq}(x,t) \\
       2n_f P_{qg}(x,t) & P_{gg}(x,t) \\
\end{array} \right)
 \otimes
 \left( \begin{array}{c}
  D_q^h(x,t)\\
  D_g^h(x,t)\\
\end{array} \right)\right]\,,
\ee
where $\otimes$ denotes as usually the convolution
$[f\otimes g]\,(z)\equiv\int_z^1 dx/x f(x)g(x/z)=\\
\int_z^1 dx/x f(x/z)g(x)$, and $P_{ij}\equiv P_{i\to j}$.

A formal extension of the standard DGLAP equations (\ref{matrix1}) 
to the SUSY case is straightforward. Denoting squark and gluino by   
$\tilde{q}$ and $\tilde{g}$, respectively, and considering 
the flavor singlet FF for squarks as it was discussed for quarks, 
we can write the SUSY DGLAP equations as
\be \label{matrix2}
\partial_{t}\left(\begin{array}{c}
 D_q^h(x,t)\\
 D_g^h(x,t)\\
 D_{\tilde{q}}^h(x,t)\\
 D_{\tilde{g}}^h(x,t)\\
\end{array} \right)=
\left[
\left( \begin{array}{cccc}
 P_{qq}(x,t) & P_{gq}(x,t) & P_{\tilde{q}q}(x,t) & P_{\tilde{g}q}(x,t) \\
 2n_f P_{qg}(x,t) &     P_{gg}(x,t) & 2n_f P_{\tilde{q}g}(x,t) & 
 P_{\tilde{g}g}(x,t) \\
 P_{q\tilde{q}}(x,t) & P_{g\tilde{q}}(x,t) & P_{\tilde{q}\tilde{q}}(x,t) &
 P_{\tilde{g}\tilde{q}}(x,t) \\
 2n_f P_{q\tilde{g}}(x,t) & P_{g\tilde{g}}(x,t) & 2n_f P_{\tilde{q}\tilde{g}}(x,t) &
 P_{\tilde{g}\tilde{g}}(x,t) \\
\end{array} \right) \otimes
\left (\begin{array}{c}
 D_q^h(x,t)\\
 D_g^h(x,t)\\
 D_{\tilde{q}}^h(x,t)\\
 D_{\tilde{g}}^h(x,t)\\
\end{array}\right) \right]\,.
\ee 
The exact form of the splitting functions $P_{ij}(x,t)$ is not
known, but using the perturbative expansion of these quantities one has
\be\label{tot_split}
 P_{ij}(x,t)=\sum_{n=0}^{\infty}
 \left(\frac{\alpha_s(t)}{2\pi}\right)^{n+1} P^{(n)}_{ij}(x) =
 \frac{\alpha_s(t)}{2\pi}\: P^{(0)}_{ij}(x) + 
 {\cal O}(\alpha_s^2) \,.
\ee
The LO splitting functions of SUSY QCD were derived in
Ref. \cite{SUSY} and are given in their unregularized form $\hat
P_{ij}$ in Table \ref{sf}. 

In the case of the diagonal splitting functions, we have to
distinguish between regularized splitting functions $P_{ii}(x)$ and
unregularized  ones $\hat P_{ii}(x)$. These splitting
functions have a probabilistic interpretation only for $x<1$, since
they contain a delta function contribution at $x=1$ accounting for
losses. Moreover, they describe the emission of soft gluons for $x\to 1$
and contain therefore a pole of infrared type which needs to be 
regularized. Using momentum conservation, 
\begin{subequations}
\ba
 && \int_0^1 dz \: z\,
 [P^{(0)}_{qq}(z)+P^{(0)}_{gq}(z)+P^{(0)}_{\tilde{q}q}(z)+
 P^{(0)}_{\tilde{g}q}(z)]=1
\\
 && \int_0^1 dz \: z\,
 [2n_fP^{(0)}_{qg}(z)+P^{(0)}_{gg}(z)+2n_fP^{(0)}_{\tilde{q}g}(z)+
 P^{(0)}_{\tilde{g}g}(z)]=1
\\
 && \int_0^1 dz \: z\,
 [P^{(0)}_{q\tilde{q}}(z)+P^{(0)}_{g\tilde{q}}(z)+
 P^{(0)}_{\tilde{q}\tilde{q}}(z)+P^{(0)}_{\tilde{g}\tilde{q}}(z)]=1
\\
 && \int_0^1 dz \: z\,
 [2n_fP^{(0)}_{q\tilde{g}}(z)+P^{(0)}_{g\tilde{g}}(z)+
 2n_fP^{(0)}_{\tilde{q}\tilde{g}}(z)+P^{(0)}_{\tilde{g}\tilde{g}}(z)]=1 \,,
\ea
\end{subequations}
one obtains as formal expression for the regularized splitting
functions 
\begin{subequations}
\ba       \label{ee1}
 && \hskip-1.6cm 
 P^{(0)}_{qq}(x) = \hat P_{qq}^{(0)}(x) - 
 \delta(1-x)\int_0^1 \!\!\! dz \: z\, 
 [\hat P^{(0)}_{qq}(z)+P^{(0)}_{gq}(z)+P^{(0)}_{\tilde{q}q}(z)+
 P^{(0)}_{\tilde{g}q}(z)] 
\\
 && \hskip-1.6cm
 P^{(0)}_{gg}(x) = \hat P^{(0)}_{gg}(x)-
 \delta(1-x)\int_0^1 \!\!\! dz \: z\, 
 [2n_f\hat P^{(0)}_{qg}(z)+P^{(0)}_{gg}(z)+2n_fP^{(0)}_{\tilde{q}g}(z)+
 P^{(0)}_{\tilde{g}g}(z)] 
\\
 &&  \hskip-1.6cm
 P^{(0)}_{\tilde{q}\tilde{q}}(x) = \hat P^{(0)}_{\tilde{q}\tilde{q}}(x)-
 \delta(1-x)\int_0^1 \!\!\! dz \: z\, 
 [\hat P^{(0)}_{q\tilde{q}}(z)+P^{(0)}_{g\tilde{q}}(z)+
 P^{(0)}_{\tilde{q}\tilde{q}}(z)+P^{(0)}_{\tilde{g}\tilde{q}}(z)]
\\  &&  \hskip-1.6cm \label{ee4}
 P^{(0)}_{\tilde{g}\tilde{g}}(x) = \hat P^{(0)}_{\tilde{g}\tilde{g}}(x)-
 \delta(1-x)\int_0^1 \!\!\! dz \: z\, 
 [2n_f\hat P^{(0)}_{q\tilde{g}}(z)+P^{(0)}_{g\tilde{g}}(z)+
 2n_fP^{(0)}_{\tilde{q}\tilde{g}}(z)+P^{(0)}_{\tilde{g}\tilde{g}}(z)]
 \,. 
\ea
\end{subequations}

Instead of calculating explicitly the expressions after the delta
functions, we substitute Eqs. (\ref{ee1}--\ref{ee4}) directly into
the DGLAP equations. In the case of ordinary QCD, the evolution of the
singlet quark FF is then given by
\ba
 \partial_{t} D_q^h(x,t) & = & 
 \frac{\alpha_s(t)}{2\pi} \int_0^1 dz \:
 \Bigg\{ \hat P^{(0)}_{qq}(z) \left [
         \frac{1}{z}D_q^h\left (\frac{x}{z},t\right )
         \Theta(z-x)- z D_q^h(x,t)\right ] - 
\nonumber\\ &&
  z P^{(0)}_{gq}(z)D_q^h(x,t) \Bigg\}+
 \frac{\alpha_s(t)}{2\pi} \int_x^1 \frac{dz}{z} \:
 P^{(0)}_{gq}(z) D_g^h\left (\frac{x}{z},t\right ) \,,
\label{equ}
\ea
where $\Theta$ denotes the usual step function. Since the two terms in 
the square bracket cancel each other for $z=1$, the pole of 
$\hat P^{(0)}_{qq}(z)$ disappears. The same method can be used to
replace the other diagonal splitting functions by their unregularized
counter-parts. 

Energy conservation is automatically ensured by Eqs. (\ref{ee1}--\ref{ee4}). 

The DGLAP equations allow to evolve the FFs $D_i^h(x)$ known at some scale 
$s_0$ to higher scales $s$. Since we are interested in a comparison
of this method with our MC simulation, we use for the initial scale 
$s_0=M_Z^2$, i.e. the same as we have used in the MC to derive the 
hadronization function. Alternatively, we shall use for the
initial scale of evolution also the very low value 
$\sqrt s_0 \sim 1$~GeV as it will be explained
in Section \ref{had-mc}.

In the case of the initial scale $M_Z$ one has two initial FFs, 
$D_q^h(x,M_Z)$ and $D_g^h(x,M_Z)$. The former is known experimentally
and the latter is calculated using our MC simulation.
We shall describe now shortly some technical aspects connected with the
numerical solution of the DGLAP equations. Starting from an initial
set of FFs at given $t$ and $x$, all
FFs are evolved simultaneously with a 4th order Runge-Kutta algorithm
with fixed step-size. At each Runge-Kutta step, the rhs of the DGLAP
equations is evaluated with a Gaussian quadrature algorithm. Since
this algorithm requires the knowledge of the FFs at
 $x$ values different from the initially chosen ones, a polynomial
interpolation algorithm  is used to calculate the FFs.
To avoid the $1/z$ singularities in the integrand, we evolve $xD_i^h(x,t)$
instead of $D_i^h(x,t)$. 

For the evolution of $\alpha_s(s)$ as a function of $s$ we use the
same method as in our MC simulation \cite{BKMC}: we combine the thresholds of
gluinos and all squarks in a single threshold at $s= M_{\rm SUSY}^2$. The
numerical value of $M_{\rm SUSY}$ is then fixed by requiring unification of
coupling constants as in the minimal SUSY SU(5) model. This simplified 
treatment of the SUSY mass spectrum allows to compare the two methods 
using the same assumptions. Since moreover the results depend only weakly 
on $M_{\rm SUSY}$, this simplification is physically reasonable. 

We shall finish this section with a remark on the connection between
the more often discussed space-like evolution of structure functions
$f$ and the time-like evolution of fragmentation functions $D$ describing
(SUSY) QCD cascades \cite{Al81}. 
Since $D_j$ represents the fragmentation of the final parton $j$,
while $f_i$ describes the distribution of the initial parton $i$,
the matrix of splitting functions $P$ has to be transposed going from
one case to the other, as 
\be \label{t1}
 \left( P_{i\bb j} \right)_{ij, \rm space-like} \leftrightarrow
 \left( P_{j\to i} \right)_{ji, \rm time-like} \,.
\ee

On the other hand, a formal analytic continuation  relates the
splitting functions in both regions at leading order (LO): neglecting
color factors, this relation is 
\be  \label{t2}
 \left|\left[ xP_{i\bb j}(1/x) \right]_{\rm space-like} \right| = 
 \left|\left[ P_{i\to j}(x) \right]_{\rm time-like} \right| \,.
\ee
Performing both transformations, the DGLAP equations are 
identical in the time- and space-like region at LO. 

Since in  Refs. \cite{SaTo,IbTo} only the transformation (\ref{t1}) 
has been performed, the splitting functions there (most notably for
gluons and gluinos) are different from ours and from those in 
Refs. \cite{BaDr1,BaDr2}, where both transformations (\ref{t1}) and  
(\ref{t2}) have been correctly used. 

\section{Hadronization functions in the Monte Carlo simulation}
\label{had-mc}
In this Section we discuss the general properties of the hadronization
functions used in the MC simulation and their connection with the
DGLAP method.  

In the MC simulation the hadronization functions are defined by
Eq.~(\ref{hfunc}). They can be determined from the FFs
at lower scales, e.g at $M_X= M_Z$, known from $e^+e^-\to$~hadrons 
data. Namely, we take the measured all-hadron spectrum as FF 
$D_q^h(x,M_Z)$ as lhs of Eq.~(\ref{hfunc}) and compute
the hadronization functions at the rhs of this equation. 

However, hadronization functions cannot be calculated in
an unique way using Eq.~(\ref{hfunc}):
Only if the FF function $D_q^h(x,M_Z)$ were known precisely at an
infinite number of points $x$, then the hadronization functions
$f_j(x)$ could be calculated in principle in an unique way using e.g. the
method of inversion. In practice  the number of points $x$, 
where the FFs are measured is limited, and the experimental errors
at some of them (most notably at $x$ close to 1) are large. 
Another uncertainty arises if the flavor dependence of FFs is considered.
The arbitrary choice of the
minimal virtuality $Q_0$ of partons in the MC also introduces some
additional error: Although the rhs of Eq. (\ref{hfunc}) should not depend on
$Q_0$, in practice all MC simulations yield slightly different results
for different $Q_0$. 

Instead of using the inversion method, we assume a specific
functional form of the hadronization function, characterized by a set
of free parameters, and perform then a fit to the data. In this method 
additional uncertainties appear, because the functional from has to be
specified a priori.
As a result one can obtain different hadronization functions within
the uncertainties discussed above. Deriving a new set of hadronization
functions in this paper, we estimate as by-product the corresponding
uncertainties in the calculated spectra. 

In Ref. \cite{BKMC} we have used two hadronization functions---one for
the quark flavor singlet and another for the gluons---motivated by the 
limiting spectrum. Each of them contains three free parameters and we
have determined these parameters using the observed spectrum of
charged hadrons at $\sqrt s=M_Z$. Then we performed several tests to
check our hadronization scheme (see Ref.~\cite{BKMC}). Among them are
the calculation of hadron spectra at two other scales, $\sqrt s=58$ GeV and 
$\sqrt s=133$ GeV, where measurements are available. We also
mention here one particular test based on the limiting spectrum. 
We run our ordinary QCD MC with the hadronization functions fixed 
as above using the same set of assumptions which are used in the
derivation of the limiting spectrum: we fixed the number of flavors to
$n_f=3$  (or to $n_f=6$) and the running of $\alpha_s$ was taken
exactly as in the limiting spectrum method. We obtained an excellent
agreement at all scales $M_X$ and for all $x$ except those close to one,
where the limiting spectrum is invalid. 

The FFs in these calculations, $D_i^h(x,M_X)$ and $D_i^j(x,M_X)$, 
conserve momentum exactly and with good accuracy, respectively.

In this Section we shall use another choice of hadronization
functions, imposing low-energy data and some additional physical 
interpretation motivated by the DGLAP equations, and demonstrate 
that the calculated spectra for different $M_X$ agree well with our
previous calculations. 

We shall begin with the general properties of fragmentation and  
hadronization functions, which we impose on those used in 
our calculations.\\ 
{\em (i) Momentum conservation of hadrons,\\
$\sum_h\int_0^1 D_i^h(x,s)xdx=1$, and partons
$\sum_j\int_0^1 D_i^j(x,s)xdx=1$, results in the normalization of the 
hadronization function as $\sum_h\int_0^1 f_i^h(x)xdx=1$.}

The proof follows from Eq. (\ref{hfunc}). Indeed,
\be
1=\sum_h\int_0^1D_i^h(x,s)xdx=\sum_h\sum_j\int_0^1xdx\int_x^1\frac{dz}{z}
D_i^j(\frac{x}{z},s)f_j^h(z) \,.
\label{norm1}
\ee

Changing the order of integration, using the variable $x'=x/z$ and
introducing the new function 
\be
\zeta_i^j(s) \equiv \int_0^1 D_i^j(x',s)x'dx',
\label{zeta}
\ee
one obtains
\be
1=\sum_h\sum_j \int_0^1 dz z f_j^h(z)\zeta _i^j(s),
\ee
which must hold for arbitrary $s$. Consistency with the condition 
$\sum_j\zeta_i^j(s)=1$ leads to
\be
\sum_h\int_0^1 dz z f_i^h(z)=1 
\label{hnorm}
\ee
as the only solution. 

Similarly, one can prove the reversed statement:\\
{\it (ii) The momentum conservation of partons in the perturbatively
calculated cascade\\  
$\sum_j\int_0^1 D_i^j(x,s)xdx=1$, with the
hadronization function normalized as in Eq.(\ref{hnorm}) results in 
momentum conservation of hadrons}.

\begin{figure}
\begin{picture}(15,9)
\put(2.5,8.3) { \epsfig{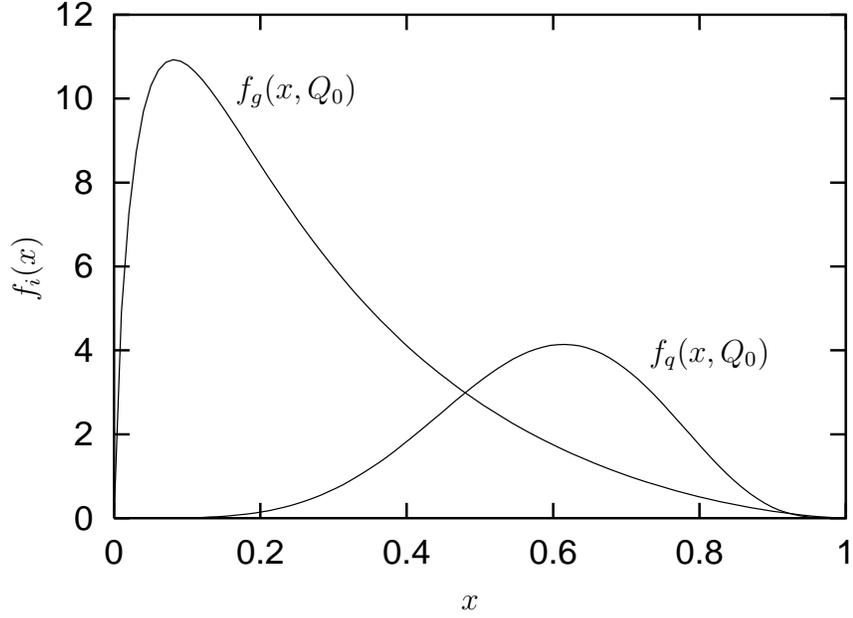} }
\put(8.0,0.2) {$x$}
\put(2.3,4.6){
\makebox(0,0)[b]{\shortstack{$f_i(x)$}}%
}
\put(5.0,7.0) {$f_g(x,Q_0)$}
\put(10.5,3.5) {$f_q(x,Q_0)$}
\end{picture}
\caption{\label{f_had}
Hadronization functions for quarks $f_q(x,Q_0)$ and
gluons $f_g(x,Q_0)$ obtained by  fitting experimental
data at $\sqrt{s}=91.2$ GeV and using $Q_0^2=0.625$ GeV$^2$.}
\end{figure}

Now we add an assumption about hadronization functions which links MC
and DGLAP methods, namely we assume that the hadronization function is 
a FF function $D_i^h(x,\sqrt s)$ extrapolated to the very low scale 
$\sqrt s=Q_0$. Then the FF $D_i^h(x,\sqrt s)$ can be 
computed with the help of DGLAP evolution equations using 
the hadronization function $f_i^h(x,Q_0)$ as the initial FF.
Under this assumption we should choose the hadronization function as 
the fragmentation function obtained for low energy $e^+e^-$
annihilation and/or $ep$ scattering. Namely, we take it following
Ref.~\cite{BKK95} as
\be
 f_i^h(x,Q_0)=N_i x^{a_i} (x+x_i)^{-b_i} (1-x)^{c_i} \,,
\label{HF}
\ee
where the index $i$ runs through quark singlet $q$ and gluon $g$, and
the index $h$ refers to the total hadron spectra. We fix the value of $Q_0$ 
in our MC simulation as in Ref.~\cite{BKMC} to $Q_0^2=0.625$ GeV$^2$.
As in \cite{BKMC}, the parameters are found performing a $\chi^2$ fit
of the LEP spectrum at $\sqrt s=M_Z$ by $D_i^h(x,M_Z)$ 
calculated from Eq. (\ref{hfunc}). The hadronization functions are 
shown in Fig. \ref{f_had}. Their main features agree with 
those of \cite{BKMC}: $f_g$ has its maximum at $x\sim 0.1$, while
$f_q$ peaks close to $x=1$. As a  difference one can see that the new
fit function is chosen to ensure $f_q(x=1)=0$.

\begin{figure}
\begin{picture}(15,9)
\put(2.5,8.3) { \epsfig{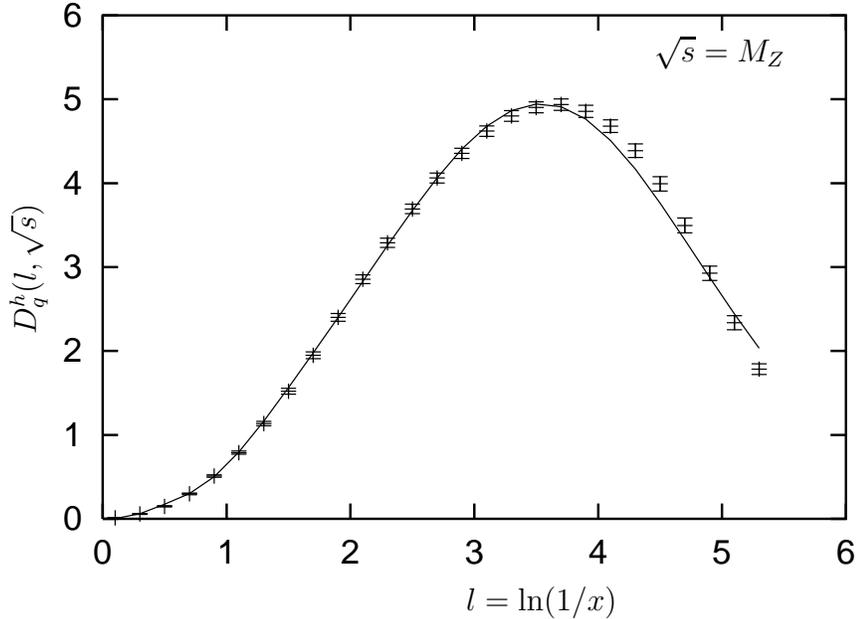} }
\put(8.0,0.2) {$l=\ln(1/x)$}
\put(2.3,4.6){
\makebox(0,0)[b]{\shortstack{$D_q^h(l,\sqrt{s})$}}%
}
\put(10.5,7.5) {$\sqrt{s}=M_Z$}
\end{picture}

\caption{\label{fitMZ}
Comparison of  the experimental data at $\sqrt{s}=91.2$ GeV
and FF $D_q^h(x,M_Z)$ computed by MC with hadronization functions as
shown in Fig.\ref{f_had}.}
\end{figure}

\begin{figure}
\begin{picture}(15,9)
\put(2.5,8.3) { \epsfig{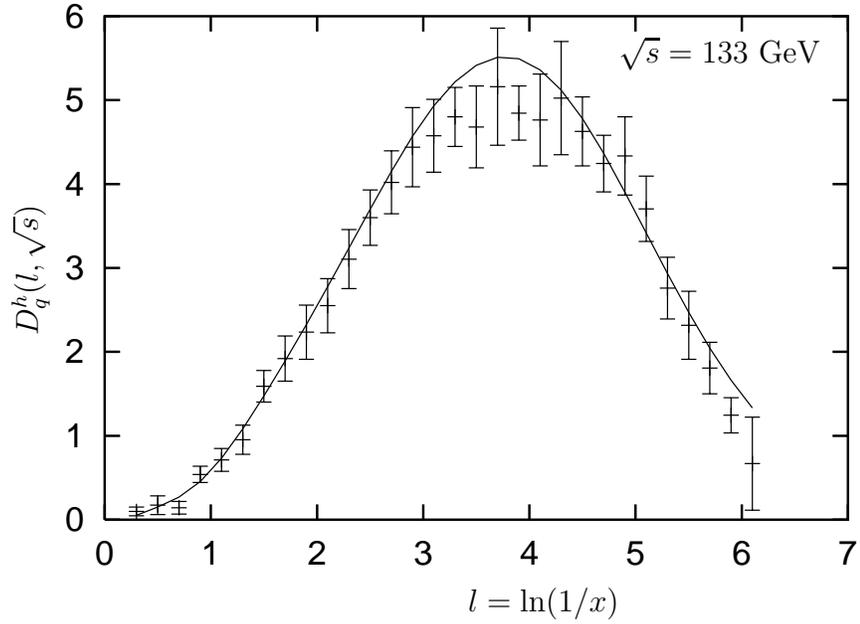} }
\put(8.0,0.2) {$l=\ln(1/x)$}
\put(2.3,4.6){
\makebox(0,0)[b]{\shortstack{$D_q^h(l,\sqrt{s})$}}%
}
\put(10.0,7.5) {$\sqrt{s}=133$ GeV}
\end{picture}
\caption{\label{fit133}
Comparison of MC computed FF $D_q^h(x,\sqrt s)$ with experimental data at
$\sqrt{s}=133$ GeV.}
\end{figure}

\begin{figure}
\begin{picture}(15,9)
\put(2.5,8.3) { \epsfig{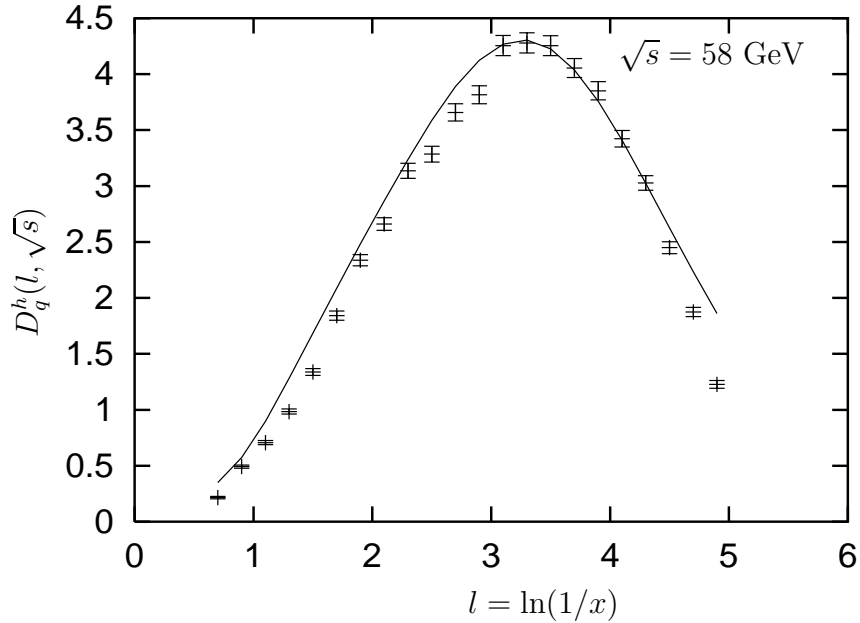} }
\put(8.0,0.2) {$l=\ln(1/x)$}
\put(2.3,4.6){
\makebox(0,0)[b]{\shortstack{$D_q^h(l,\sqrt{s})$}}%
}
\put(10.0,7.5) {$\sqrt{s}=58$ GeV}
\end{picture}
\caption{\label{fit58}
Comparison of MC computed FF $D_q^h(x,\sqrt s)$ with experimental data at
$\sqrt{s}=58$ GeV.}
\end{figure}

\begin{figure}
\begin{picture}(15,9)
\put(2.5,8.3) { \epsfig{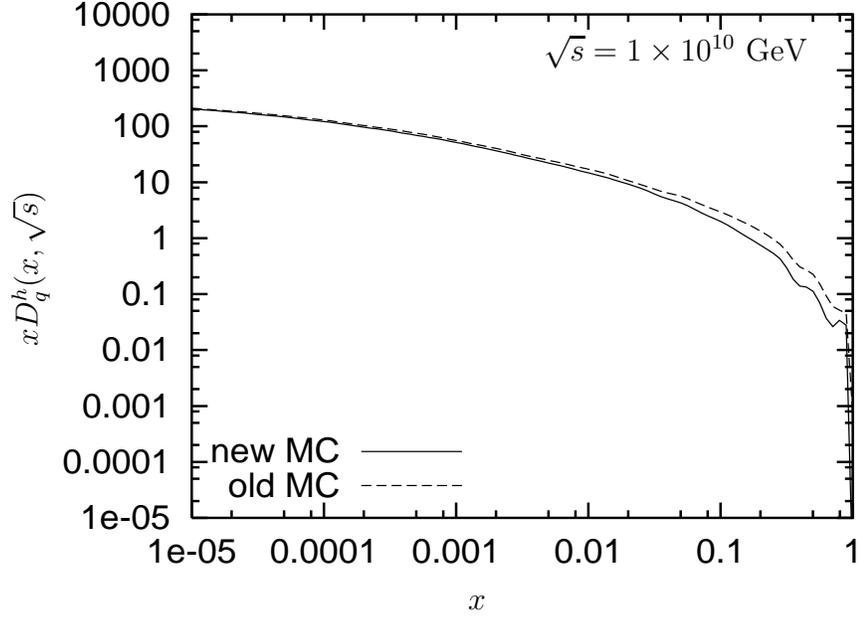} }
\put(8.0,0.2) {$x$}
\put(2.3,4.6){
\makebox(0,0)[b]{\shortstack{$xD_q^h(x,\sqrt{s})$}}%
}
\put(9.0,7.5) {$\sqrt{s}=1\times 10^{10}$ GeV}
\end{picture}
\caption{\label{MCfh1}
Comparison of spectra calculated with the old \cite{BKMC} and
new hadronization
functions: FFs from  ordinary
QCD MC with quark as a primary parton are displayed for the scale
$\sqrt s=1\times 10^{10}$ GeV for the new (solid line) and old (dashed line) 
hadronization functions.}
\end{figure}

\begin{figure}
\begin{picture}(15,9)
\put(2.5,8.3) { \epsfig{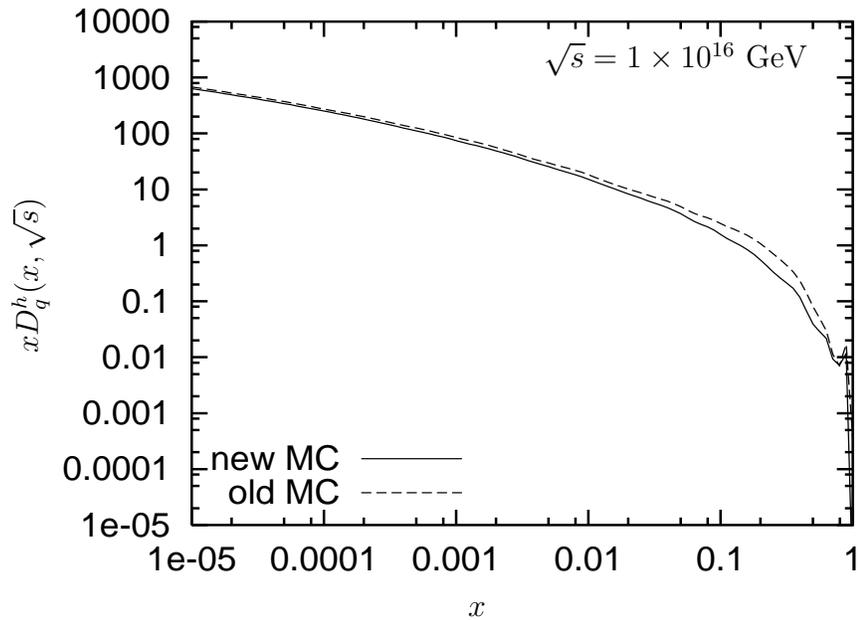} }
\put(8.0,0.2) {$x$}
\put(2.3,4.6){
\makebox(0,0)[b]{\shortstack{$xD_q^h(x,\sqrt{s})$}}%
}
\put(9.0,7.5) {$\sqrt{s}=1\times 10^{16}$ GeV}
\end{picture}
\caption{\label{MCfh2}
The same as in Fig. \ref{MCfh1} for the scale $\sqrt s=1\times 10^{16}$ GeV.}
\end{figure}

\begin{figure}
\begin{picture}(15,9)
\put(2.5,8.3) { \epsfig{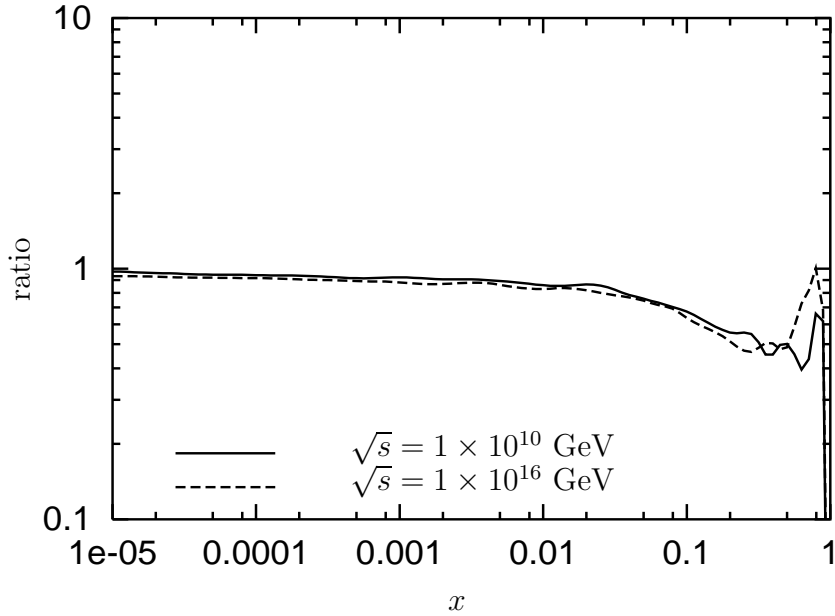} }
\put(8.0,0.2) {$x$}
\put(2.3,4.6){
\makebox(0,0)[b]{\shortstack{ratio}}%
}
\put(6.7,2.25) {$\sqrt{s}=1\times 10^{10}$ GeV}
\put(6.7,1.80) {$\sqrt{s}=1\times 10^{16}$ GeV}
\end{picture}
\caption{\label{ratio1}
Ratio of the spectra calculated with the old and new hadronization functions
for the scale $\sqrt s=1\times 10^{10}$~GeV (solid line) and 
$\sqrt s=1\times 10^{16}$~GeV (dashed line). For discussion of
uncertainties at large $x > 0.1$ see the text. 
}
\end{figure}

In Figs. \ref{fitMZ}, \ref{fit133} and \ref{fit58}, the spectra of hadrons 
$D_q^h(x,\sqrt s)$ calculated for the scales $\sqrt s= 58$~GeV,
$91.2$~GeV, and $133$~GeV are compared with observations. (The
measurements are for charged hadrons and rescaled by us to 
the total hadron spectra.)
In Figs. \ref{MCfh1} and \ref{MCfh2} these  spectra computed at the scales
$\sqrt s = 1\times 10^{10}$ GeV and $1\times 10^{16}$ GeV using the
old \cite{BKMC} and new hadronization functions are shown. The
agreement is good, as it has been expected. The ratio between the new and 
old spectra, shown in Fig. \ref{ratio1}, illustrates the
uncertainties in our MC simulations connected with 
different choices of hadronization functions. Figures \ref{MCfh1}, 
\ref{MCfh2} and \ref{ratio1} show that these uncertainties affect only 
the high energy part of the spectrum at $x>0.1$.

We described above the all-hadron spectra $D_i^h(x,M_X)$. Similar
calculations have been performed by us for charged pions and nucleons,  
using the experimental data at $\sqrt{s}=M_Z$ from the ALEPH and OPAL 
collaborations~\cite{ALEPH-OPAL}.

As a test of the interpretation of the MC hadronization function 
$f_i^h(x,Q_0)$ as low energy limit of the FF $D_i^h(x,\sqrt s)$,
we have evolved the hadronization function given by Eq.~(\ref{HF}) from 
the scale $Q_0^2=0.625$ GeV$^2$ to the higher scales using our QCD DGLAP
code described in Section~\ref{DGLAP_sec}.
In principle, this procedure should not work well for very small 
$x=2p/Q_0$, where one is physically in the non-perturbative regime 
(see Introduction). But using the DGLAP equations which prescribe $s$
as argument of $\alpha_s$, we can perform formally the evolution calculations,
as it was done in Refs.~\cite{Kniehl,BKK95} and \cite{BaDr1,BaDr2}.

\begin{figure}
\begin{picture}(15,9)
\put(2.5,8.3) { \epsfig{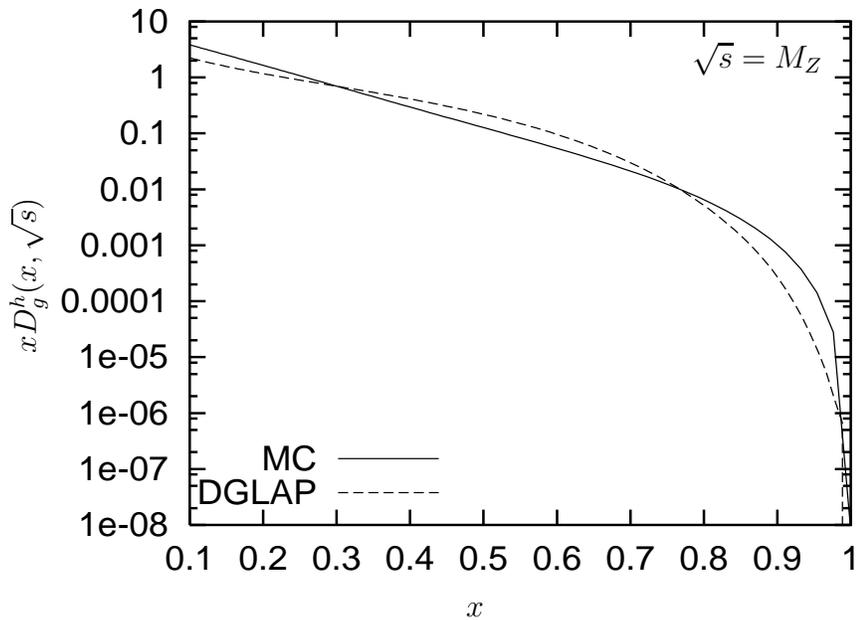} }
\put(8.0,0.2) {$x$}
\put(2.3,4.6){
\makebox(0,0)[b]{\shortstack{$xD_g^h(x,\sqrt{s})$}}%
}
\put(11.0,7.5) {$\sqrt{s}=M_Z$}
\end{picture}
\caption{\label{fhevol1}
DGLAP FF with hadronization functions as initial fragmentation functions
at the scale $Q_0$.  
The FF $D_g^h(x,\sqrt s)$ is DGLAP evolved from
the scale $\sqrt s=Q_0$ to the scale $M_Z$ (dashed curve) and plotted 
in comparison with MC calculated $D_g^h(x,M_Z)$ (full curve).}
\end{figure}

\begin{figure}
\begin{picture}(15,9)
\put(2.5,8.3) { \epsfig{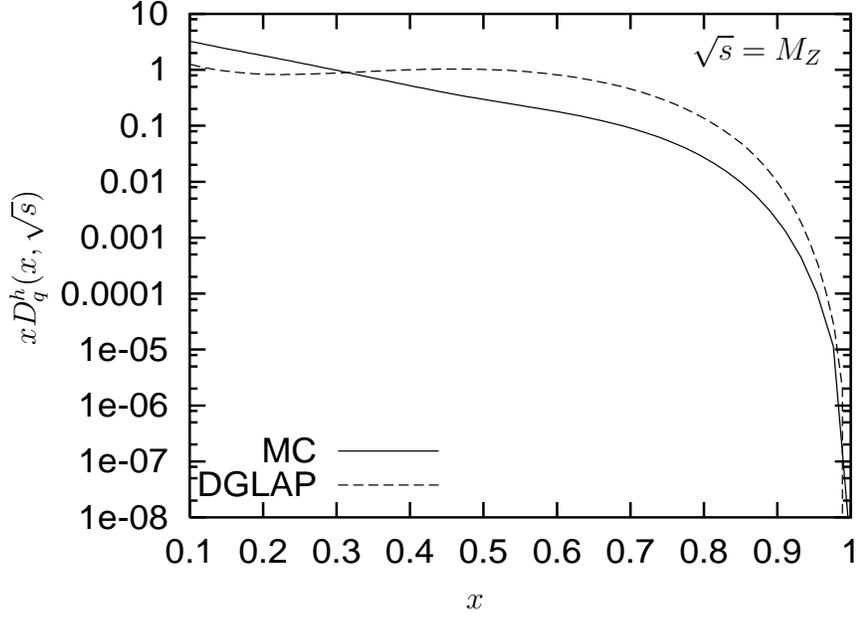} }
\put(8.0,0.2) {$x$}
\put(2.3,4.6){
\makebox(0,0)[b]{\shortstack{$xD_q^h(x,\sqrt{s})$}}%
}
\put(11.0,7.5) {$\sqrt{s}=M_Z$}
\end{picture}
\caption{\label{fhevol2}
The same as in Fig. \ref{fhevol1} for the quark FF $D_q^h(x,M_Z)$.}
\end{figure}

\begin{figure}
\begin{picture}(15,9)
\put(2.5,8.3) { \epsfig{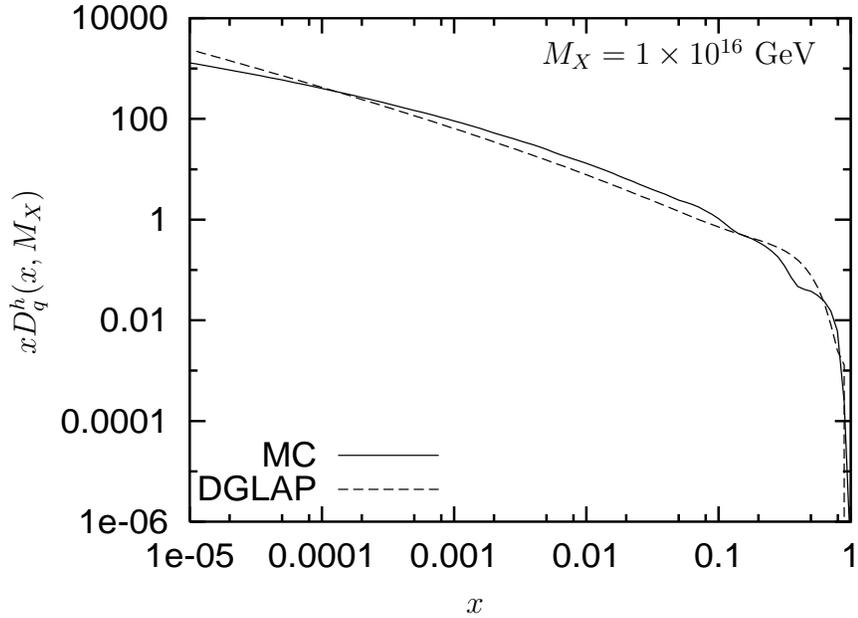} }
\put(8.0,0.2) {$x$}
\put(2.3,4.6){
\makebox(0,0)[b]{\shortstack{$xD_q^h(x,M_X)$}}%
}
\put(9.0,7.5) {$M_X=1\times 10^{16}$ GeV}
\end{picture}
\caption{\label{fhevol3}
SUSY DGLAP FF with hadronization functions as initial fragmentation functions
at the scale $Q_0$. The FF $D_q^h(x,M_X)$ is SUSY DGLAP evolved from
the scale $\sqrt s=Q_0$ to scale $M_X=10^{16}$ GeV (dashed curve) and plotted 
in comparison with MC calculated $D_q^h(x,M_Z)$ (full curve).}
\end{figure}

In Fig. \ref{fhevol1} and \ref{fhevol2} the evolved hadronization 
functions at the scale $M_Z$ are compared with FFs $D_i^h(x,M_Z)$ 
calculated with the MC. We find good agreement for $D_g^h$ but much worse 
agreement for $D_q^h$. We think that a reason for this failure is the
use of the quark flavor singlet FF at low scales. Indeed, in the
successful evolution  
of the Refs.~\cite{BaDr1,BaDr2,Kniehl,BKK95}, the FFs for different 
quarks have different 
parameterizations. For scales above $M_Z$, the flavor singlet FF becomes
a good approximation, and, indeed, the evolution of hadronization
functions from the scale $Q_0$ to scale $M_{\rm GUT}$ results in very good 
agreement with MC simulation (see Fig. \ref{fhevol3}). 
In this case, the agreement is equally good for both $D_q^h$ and  $D_g^h$.

\section{Comparison of DGLAP and MC hadron spectra}
\label{comparison}
In this Section we shall compare the hadron spectra computed by
the two methods discussed above, MC and DGLAP. 

We shall begin with a remark concerning the smallest and largest $x$
values of practical interest. Coherent branching produces the
so-called Gaussian peak in multiplicity with maximum at
$x_{\max} \sim (Q_0/M_X)^{0.6} \sim 2\times 10^{-10}$ for 
$M_X \sim M_{\rm GUT}$.
This approximate estimate coincides well with the value found in the
MC simulation~\cite{BKMC}. Note that the DGLAP method is not valid for
those $x$ values where coherence plays an essential role, namely at $x \lsim
4\times 10^{-5}$ (see Fig.~8 in Ref.~\cite{BKMC}). The lowest value 
$x_{\rm min}$ relevant for physical applications  is much higher  
than $x_{\max}$: 
$x_{\rm min}\sim 2E_{\rm obs}/M_X \sim 2\times 10^{-6}$ for
the same $M_X\sim M_{\rm GUT}$ and for 
$E_{\rm obs} > 1\times 10^{10}$~GeV. On the other side, the maximal
observed energy $\sim 3\times 10^{20}$~eV and the minimal $M_X$ of
interest, $M_X \gsim 10^{12}$~GeV,
restrict $x$ to values much smaller than 0.6. Thus the value 
$x=0.6$ can be considered as the largest $x$ of interest for existing
experimental data.

\begin{figure}
\begin{picture}(15,9)
\put(2.5,8.3) { \epsfig{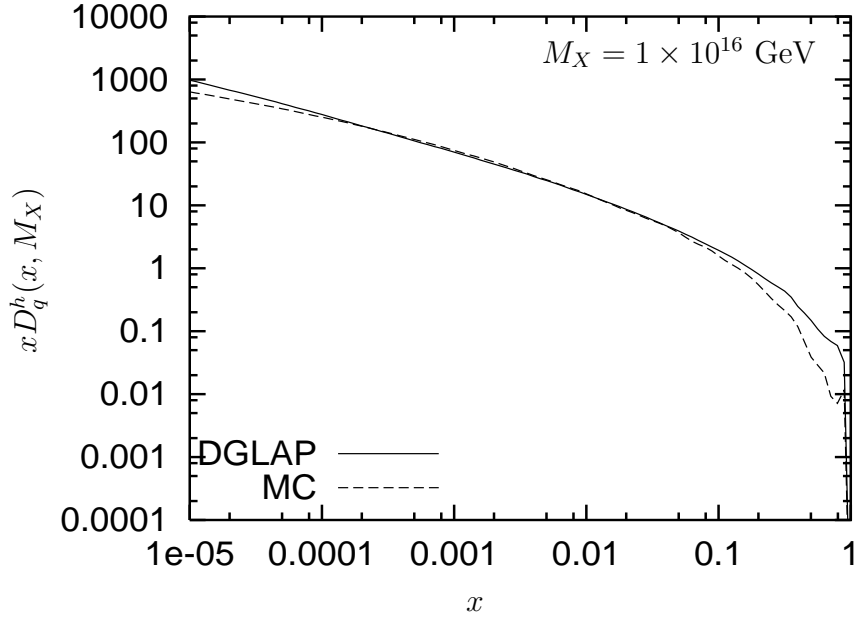} }
\put(8.0,0.2) {$x$}
\put(2.3,4.6){
\makebox(0,0)[b]{\shortstack{$xD_q^h(x,M_X)$}}%
}
\put(9.0,7.5) {$M_X=1\times 10^{16}$ GeV}
\end{picture}
\caption{\label{MC-DGLAP1}
Comparison of the DGLAP FF (solid line) and MC FF (dashed line) 
for ordinary QCD with $M_X=1\times 10^{16}$ GeV and with quark as a
primary parton.}
\end{figure}

In Fig. \ref{MC-DGLAP1} we present a comparison of $D_i^h(x,M_X)$ with
$i=q$ and $M_X=1\times 10^{16}$ GeV calculated for ordinary QCD with
the MC and DGLAP method. The initial scale in the DGLAP method is taken
as $\sqrt s=M_Z$, but the initial scale $\sqrt s=Q_0$ gives practically
an identical spectrum.
The spectra for $i=g$ agree equally well, as well as the spectra for other
high scales $M_X$.

One can see that the MC and DGLAP spectra slightly differ at very low 
$x$ and have a more pronounced disagreement at large values of $x$.
The discrepancy at low $x$ is due to coherent branching, and it starts 
at the value of $x$ estimated above ($x\lsim 4\times 10^{-5}$).
At large $x$, the calculations by 
both methods suffer from uncertainties, particularly the MC simulation. 
In this region the results are sensitive to the 
details of the hadronization scheme (see, e.g.,  the problem of
HERWIG~\cite{Rub} with the overproduction of protons at large $x$ and
the dependence on the choice of the hadronization functions in our MC
as illustrated by Figs. \ref{MCfh1} and \ref{MCfh2}).  One can add to 
this problem 
large uncertainties in the measured FFs at $M_Z$ and $x\sim 1$ and also 
the theoretical uncertainties connected with 
the models of $X$ particle decay (unknown number and types of the initial
partons\footnote{$X$ particles decay in many models due to
nonperturbative interactions, and the number of primary partons is
therefore model dependent. Since at large $x$ the number of cascade
generations is small, the FFs are for $x\to 1$ rather sensitive to the initial
multiplicity. In contrast, at small $x$ when the number of cascade
generations is large, differences in the initial multiplicity become
inessential.}  
and unknown matrix element of the $X$ particle decay). 

However, as a whole, Fig.~\ref{MC-DGLAP1} demonstrates good
agreement between MC and DGLAP methods. In Fig.~\ref{ratio2}
we present the ratio of FFs calculated with MC and DGLAP for ordinary QCD. 
At $x \leq 1\times 10^{-5}$ the MC FF is noticeably smaller than the
DGLAP FF because coherence effects suppress branchings, an effect not
included in the DGLAP method. At large $x\geq 0.1$, the discrepancy
becomes greater than 20\% due to the reasons explained above. For
$2.5\times 10^{-5}\leq x \leq 0.1$,  the agreement between these two
FFs is better than $20\%$. 

Let us now come over to SUSY QCD.

\begin{figure}
\begin{picture}(15,9)
\put(2.5,8.3) { \epsfig{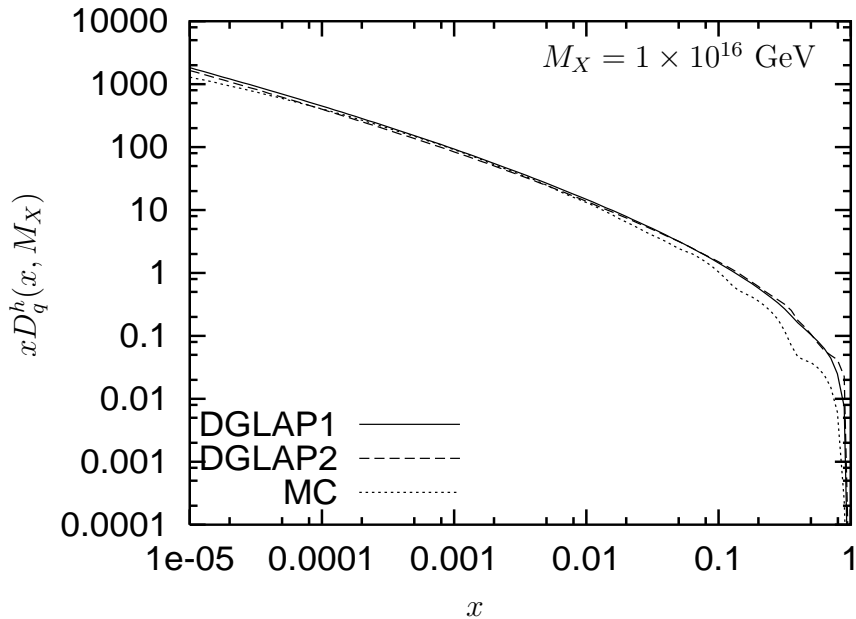} }
\put(8.0,0.2) {$x$}
\put(2.3,4.6){
\makebox(0,0)[b]{\shortstack{$xD_q^h(x,M_X)$}}%
}
\put(9.0,7.5) {$M_X=1\times 10^{16}$ GeV}
\end{picture}
\caption{\label{MC-DGLAP2}
Comparison of SUSY DGLAP and SUSY MC fragmentation functions for
$M_X=1\times 10^{16}$ GeV with quark as a primary parton.
SUSY DGLAP FFs are calculated for 10 TeV as the starting scale
(solid line) and for $M_Z$ (broken line). 
SUSY MC FF is shown by dotted line.}
\end{figure}

\begin{figure}
\begin{picture}(15,9)
\put(2.5,8.3) { \epsfig{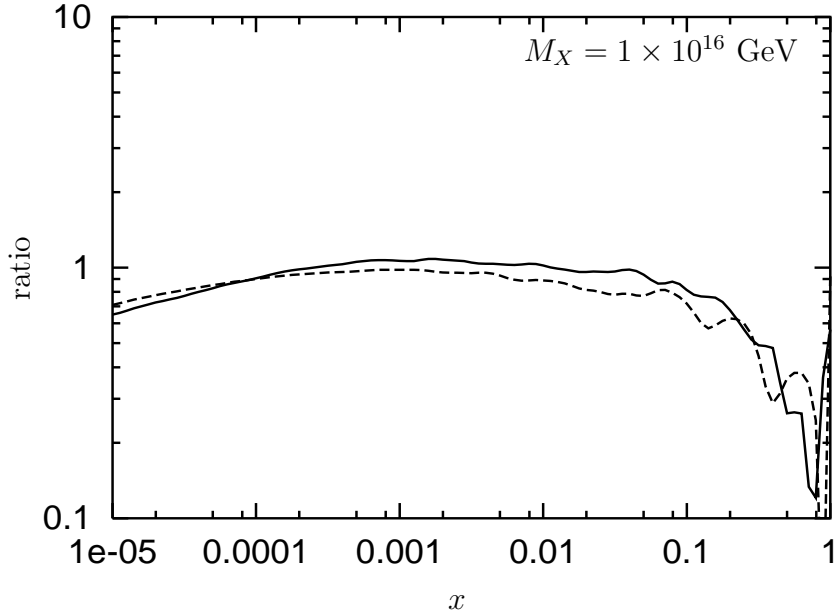} }
\put(8.0,0.2) {$x$}
\put(2.3,4.6){
\makebox(0,0)[b]{\shortstack{ratio}}%
}
\put(9.0,7.5) {$M_X=1\times 10^{16}$ GeV}
\end{picture}
\caption{\label{ratio2}
Ratio of FFs calculated by MC and DGLAP in ordinary QCD (solid curve)
and SUSY QCD (dashed curve) for $M_X=1\times 10^{16}$~GeV. For
discussion see the text.
}
\end{figure}

In Fig. \ref{MC-DGLAP2} we plot the FFs $D_i^h(x,M_X)$ calculated by MC and 
DGLAP methods for $M_X=1\times 10^{16}$ GeV and $i=q$. In the DGLAP
method the SUSY FFs have been evolved from the ones obtained with the
SUSY MC at the scale $\sqrt s= 10M_{\rm SUSY}\approx 10$~TeV. One can
see the good agreement between DGLAP (solid curve) and MC (dotted curve).
This good agreement holds also for other (lower) scales $M_X$ and for
other initial partons $i=g, \tilde g, \tilde q$. The ratio
of FFs for SUSY MC and SUSY DGLAP is shown in Fig. \ref{ratio2}.

When one does not have the initial SUSY FFs from a MC simulation, the
question arises how to proceed. As was first suggested in
Ref.~\cite{Rub,Co}, the initial FFs can be taken as the ones for ordinary
QCD at the low scale $\sqrt s=M_Z$, while the production of SUSY
partons is included in the splitting functions assuming
threshold behavior at $M_{\rm SUSY} \sim 1$~TeV. We can check this
assumption computing the SUSY FF in both ways. In Fig. \ref{MC-DGLAP2}
we present the SUSY FFs $D_i^h(x,M_X)$ for $i=q$ and $M_X=1\times
10^{16}$ GeV, evolved from the initial scale $\sqrt s=M_Z$ (dashed
curve). The good agreement between the two DGLAP curves proves the
validity of the assumption made above.

\section{Photon, neutrino and nucleon spectra}
\label{spectra}

The spectra of photons, neutrinos and nucleons produced by the decay of
superheavy particles are of practical interest in high energy
astrophysics. These spectra $D_i^a(x,M_X)$ with $a=\gamma,\nu,N$ can
be also considered as FFs. Because the dependence on the type $i$
of the primary parton is weak, we shall omit the index $i$ from now on,
keeping $a$ as subscript.

\begin{figure}
\begin{picture}(15,9)
\put(2.5,8.3) { \epsfig{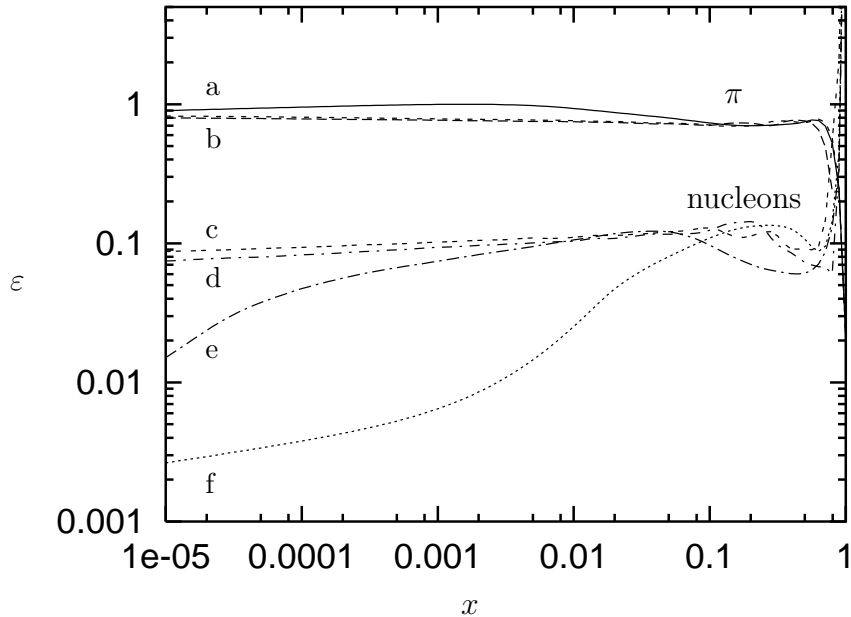} }
\put(8.0,0.2) {$x$}
\put(2.0,4.5) {$\eps$}
\put(11.0,5.6) {nucleons}
\put(11.5,7.0) {$\pi$}
\put(4.6,7.1) {\small{\small{a}}}
\put(4.6,6.4) {\small{\small{b}}}
\put(4.6,5.2) {\small{\small{c}}}
\put(4.6,4.55) {\small{\small{d}}}
\put(4.6,3.6) {\small{\small{e}}}
\put(4.6,1.8) {\small{\small{f}}}
\end{picture}
\caption{\label{eps_i}
Fraction of pions, $\epsilon_{\pi}$, and nucleons, $\epsilon_N$
relative to all hadrons according to SUSY MC simulations.
For pions the curve $a$ corresponds to $M_X=M_Z$, while curve $b$ 
describes with a good accuracy the scales $10^{10}\leq M_X \leq 10^{16}$~GeV.
For nucleons curves $c, d, e, f$ correspond to $M_X$ equal to 
$10^{16}, 10^{10}, 10^5$~GeV and $M_Z$, respectively.}
\end{figure}

Till now we concentrated our discussion on the total number
of hadrons ($a=h$) described by the FF $D_h(x,M_X)$, 
but in fact we have performed similar calculations separately for
charged pions and protons+antiprotons. The procedure of the
calculations is 
identical to that described in Section \ref{DGLAP_sec} for the DGLAP
method and in Section \ref{had-mc} for the MC. For charged pions and 
protons+antiprotons we used experimental data from Refs.~\cite{ALEPH-OPAL}.
Below we shall present results of our SUSY MC simulations in terms of 
FFs for all pions $D_{\pi}$, all nucleons $D_N$ and all hadrons $D_h$.
We introduce the ratios $\eps_N(x)$ and $\eps_{\pi}(x)$ as
\be  \label{nucl}
 D_N(x) = \eps_N(x) D_h(x)
\,,\qquad\quad
 D_{\pi}(x) = \eps_{\pi}(x) D_h(x) \,.
\ee
The spectra of pions and nucleons at large $M_X$ have approximately the
same shape as the hadron spectra, and one can use in this case
$\eps_{\pi}=0.73\pm 0.03$ and $\eps_{N}=0.12\pm 0.02$,   
taking into account the errors in the experimental 
data \cite{KEK,LEP2,ALEPH-OPAL}.
In Fig. \ref{eps_i} the ratios $\eps_N(x)$ and $\eps_{\pi}(x)$ are
plotted as functions of $x$ for different values of $M_X$. Note the 
peculiar dependence of  $\eps_N(x)$ for small $M_X$. The smallness of 
$\eps_N(x)$ at small $x$ and small $M_X$ is caused formally by the 
smallness of the observed $\pi/p$ ratio at $M_X=M_Z$, where we
fit our FFs. Physically it is due to the combined effect of coherence
and the mass difference of nucleons and pions. Indeed, the $x$ values 
at small $M_X$ where $\eps_N(x)$ is particularly small belong to the
region below the Gaussian peak ($x_m\sim (\Lambda/M_X)^{0.6}$), where
coherence effects suppress branchings particularly strong. 

We can calculate now the spectra of photons and neutrinos produced by
the decays of pions neglecting the small contribution ($0.15 \pm 0.04$)
of $K$, $D$, $\Lambda$ and other particles. Including these particles
affects stronger neutrinos than photons, which are the main topic of
this Section. For the pion spectrum we shall use
$D_{\pi}(x)=\eps_{\pi}(x)D_h(x)$.

The normalized photon spectrum from the decay of one $X$ particle at rest
is given by
\be \label{photon}
 D_{\gamma}(x) = \frac{2}{3} \int_x^1 
                        \frac{dy}{y}\:\eps_\pi(y)\:D_h(y) \,.
\ee

The total neutrino spectrum from decays of charged pions and
muons can be represented as \cite{gaisser}
\be
D_{\nu}(x)=D_{\nu_{\mu}}^{\pi\to\mu\nu_\mu}(x) +
           D_{\nu_{\mu}}^{\mu\to\nu_\mu\nu_e e}(x) +
           D_{\nu_e}^{\mu\to\nu_\mu\nu_e e}(x) \,,
\label{neutr}
\ee
where for pion decay
\be
D_{\nu_{\mu}}^{\pi\to\mu\nu_\mu}(x)=\frac{2}{3} R
\int_{xR}^1  \frac{dy}{y}\: \eps_\pi(y) D_h(y)
\label{pion}
\ee
with
\be
R= (1-m_{\mu}^2/m_{\pi}^2)^{-1},
\label{R}
\ee
and for muon decay
\be
D_{\nu_i}^{\mu\to\nu_\mu\nu_e e}(x)=\frac{2}{3}\eps_\pi R
\int_x^1 \frac{dy}{y} q_i(y) \int_{x/y}^{x/ry} 
\frac{dz}{z}~\eps_\pi(z)~ D_h(z)
\label{muon}  
\ee
with
\be
q_i= \frac{5}{3} - 3y^2+\frac{4}{3} y^3 ~ {\rm for}~~ \nu_{\mu},\bar{\nu}_{\mu}
\,\,\quad {\rm and} \,\,\quad 
q_i= 2(1 - 3y^2+ 2 y^3) ~ {\rm for}~~ \nu_e,\bar{\nu}_e \,.
\label{q}
\ee

\begin{figure}
\begin{picture}(15,9)
\put(2.5,8.3) { \epsfig{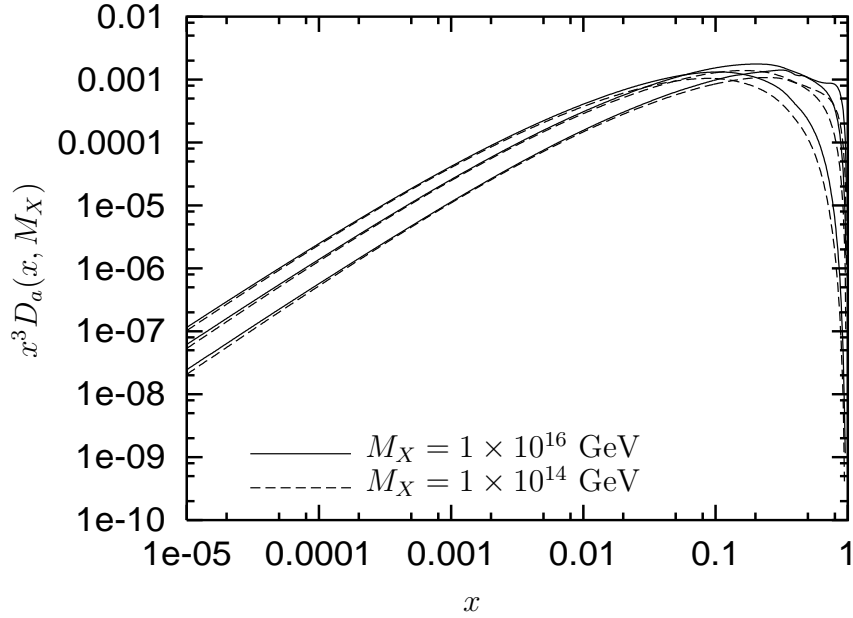} }
\put(8.0,0.2) {$x$}
\put(2.3,4.6){
\makebox(0,0)[b]{\shortstack{$x^3D_a(x,M_X)$}}%
}
\put(6.7,2.25) {$M_X=1\times 10^{16}$ GeV}
\put(6.7,1.80) {$M_X=1\times 10^{14}$ GeV}
\end{picture}
\caption{\label{Spectra}
Neutrino (upper curves), gamma (middle curves) and nucleon (lower curves)
spectra from DGLAP evolution at scales $M_X=1\times 10^{16}$ GeV
(solid line) and $M_X=1\times 10^{14}$ GeV (dashed line). The spectra
are obtained averaging over primary partons.}
\end{figure}

The spectra are presented in Fig. \ref{Spectra} for different masses $M_X$.
We shall compare our photon spectra with those calculated by the DGLAP
method in Refs.~\cite{BaDr1,BaDr2} and \cite{SaTo}. The photon spectrum is
most interesting to compare, because it is straightforwardly
related to the hadron spectrum which is the basic physical quantity.
Moreover, the photon spectrum is the dominant component of radiation
produced by superheavy particles.

\begin{figure}
\begin{picture}(15,9)
\put(2.5,8.3) { \epsfig{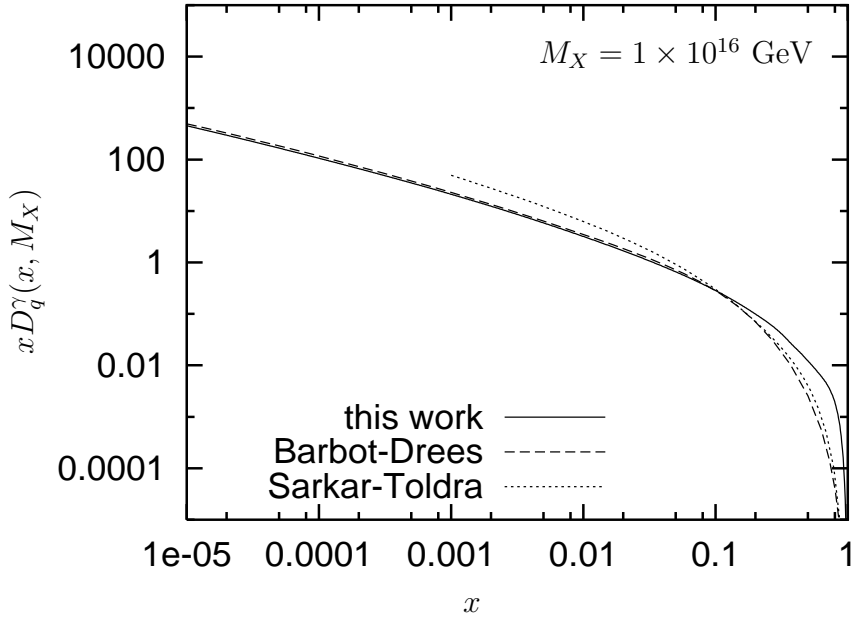} }
\put(8.0,0.2) {$x$}
\put(2.3,4.6){
\makebox(0,0)[b]{\shortstack{$xD_q^{\gamma} (x,M_X)$}}%
}
\put(9.0,7.5) {$M_X=1\times 10^{16}$ GeV}
\end{picture}
\caption{\label{comp}
Comparison of photon spectra from present work computed with 
DGLAP equation (solid line),
from \cite{BaDr1,BaDr2} (dashed line) and \cite{SaTo} (dotted line).
All three calculations are performed with quark as initial parton
for $M_X=1\times 10^{16}$ GeV.}
\end{figure}

To be precise, we compare the FF $D_q^{\gamma}(x,M_X)$ at 
$M_X=1\times 10^{16}$ GeV. Figure \ref{comp} demonstrates good 
agreement between our spectrum and those from
Refs.~\cite{BaDr1,BaDr2} at $x \leq 0.3$. 
As was it mentioned above, the disagreement at large 
$x$ is not surprising. 
Apart from $D_q^h(x,M_Z)$ taken directly from 
the experiments, both calculations use  the much more uncertain
$D_g^h(x,Q^2)$. In our case, $D_g^h(x,Q^2)$ is taken from our MC
simulation~\cite{BKMC}, in the case of Ref.~\cite{BaDr1,BaDr2} from
the fit performed in Ref.~\cite{Kniehl}. In both cases, rather large
uncertainties exist at large $x$ (e.g. see Fig.~5 from
Ref.~\cite{Kniehl}). In both calculations 
the decay of the $X$ particle into two partons is considered, but in fact 
in many models only many-parton decays exist. This and the unknown type
of the initial partons add a common
theoretical uncertainty to both calculations (see Section \ref{comparison}
for discussion)

The photon spectrum of Ref.~\cite{SaTo} shows some deviation 
at $x<0.3$ from both spectra discussed above.
To find the reason we performed the same calculations using splitting 
functions according the prescription of \cite{SaTo} (see
Section \ref{DGLAP_sec}) and
obtained indeed some excess at small $x$ and good  agreement at large $x$, 
as one observes in Fig. \ref{comp}. Nevertheless, the agreement between 
the three curves as presented in Fig. \ref{comp} is good.

\begin{figure}
\begin{picture}(15,9)
\put(2.5,8.3) { \epsfig{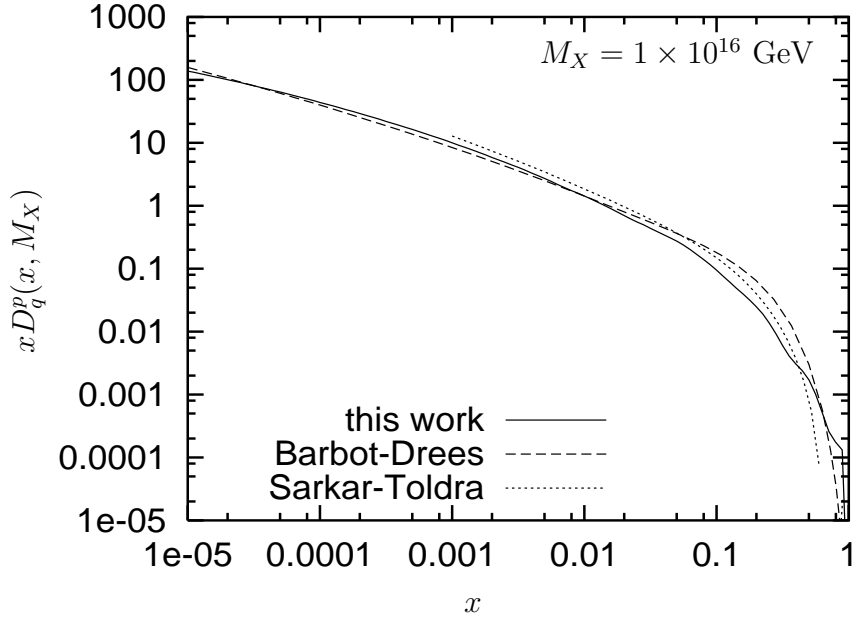} }
\put(8.0,0.2) {$x$}
\put(2.3,4.6){
\makebox(0,0)[b]{\shortstack{$xD_q^{p} (x,M_X)$}}%
}
\put(9.0,7.5) {$M_X=1\times 10^{16}$ GeV}
\end{picture}
\caption{\label{compPr}
Comparison of proton spectra from present work computed with MC (solid line),
from \cite{BaDr1,BaDr2} (dashed line) and from \cite{SaTo} (dotted line). 
All spectra are computed 
with quark as initial parton for $M_X=1\times 10^{16}$ GeV.}
\end{figure}

In Fig. \ref{compPr} we present also the comparison of our proton spectrum,
computed with the SUSY QCD MC, with that of Refs. \cite{BaDr1,BaDr2} and 
\cite{SaTo}, which shows good agreement.

\section{UHECR from Superheavy DM and Topological Defects}
\label{UHECR}

As follows from Section \ref{spectra}, the accuracy of spectrum 
calculations has reached such a level that one can consider the spectral
shape as a signature of the model. The predicted spectrum is
approximately $\propto dE/E^{1.9}$ in the region of $x$ at interest, and
it is considerably steeper than the QCD MLLA spectrum used in the end of
90s. The generation spectra for nucleons, neutrinos and gammas are
shown Fig.~\ref{Spectra}.  

\begin{figure}
\begin{picture}(15,9)
\put(2.5,8.3) { \epsfig{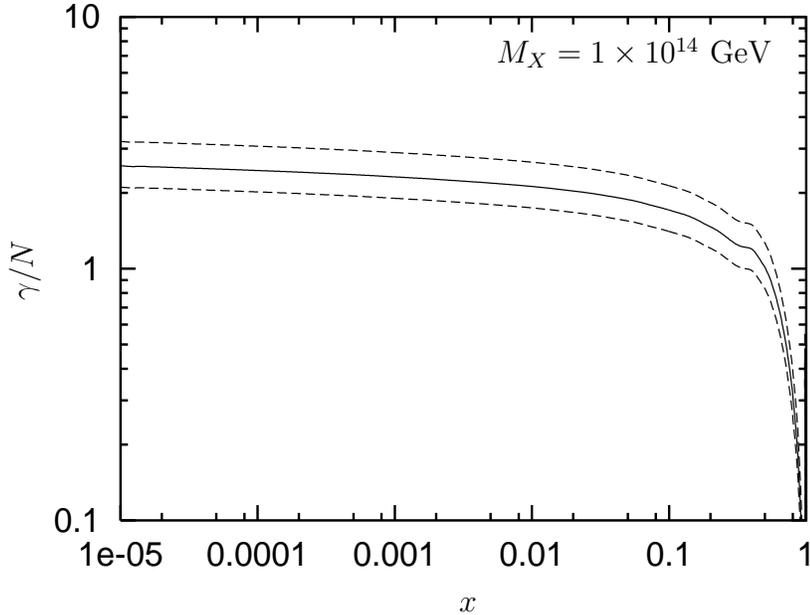} }
\put(8.0,0.2) {$x$}
\put(2.3,4.6){
\makebox(0,0)[b]{\shortstack{$\gamma / N$}}%
}
\put(8.5,7.5) {$M_X=1\times 10^{14}$ GeV}
\end{picture}
\caption{\label{ratio}
Gamma/nucleon ratio in generation spectra for $M_X=1\times 10^{14}$~GeV 
computed with MC. The dashed curves illustrate the uncertainties of 
calculations. Calculations with DGLAP give very similar results.}
\end{figure}

Another interesting feature of these new calculations is a decrease
of the ratio of photons to nucleons, $\gamma/N$, in the generation
spectrum. This ratio is presented in Fig. \ref{ratio} for $M_X=
1\times 10^{14}$ GeV by a solid curve together with a band of
uncertainties given by the two dashed curves. 
At $x\sim 1\times 10^{-3}$ this ratio is characterized by a value 
of 2 -- 3 only. The decrease of the $\gamma/N$ ratio is caused by 
a decrease of the number of pions in the new calculations and by 
an increase of 
the number of nucleons. This result has an important impact for SHDM 
and topological defect models because the fraction of nucleons
in the primary radiation increases. However, in both models photons dominate 
(i.e. their fraction becomes $\gsim 50\%$)
at $E\gsim (7-8)\times 10^{19}$~eV (see below).

In this Section we shall consider two applications: superheavy dark matter
(SHDM) and topological defects (TD).

UHECRs from SHDMs have been suggested in Refs.~\cite{BKV97,KR98}
and further studied in Ref.~\cite{Sar}. Production of SHDM particles 
naturally occurs in the time-varying gravitational field of the expanding 
universe at the post-inflationary stage~\cite{grav}.

The relic density of these particles is mainly determined
(at fixed reheating temperature and inflaton mass) by their
mass $M_X$. The range of practical interest is $(3 - 10)\times
10^{13}$ GeV, at larger masses the SHDM is a subdominant component of
the DM. 

SHDM is accumulated in the Galactic halo with the overdensity 
\be
\delta= \frac{\bar{\rho}_X^{\rm halo}}{\rho_X^{\rm extr}}=
\frac{\bar{\rho}_{\rm DM}^{\rm halo}}{\Omega_{\rm CDM}\rho_{\rm cr}} \,,
\label{overd}
\ee
where $ \bar{\rho}_{\rm DM}^{\rm halo}\approx 0.3$ GeV/cm$^3$, 
$\rho_{\rm cr}=1.88\times 10^{-29}h^2$ g/cm$^3$ and 
$\Omega_{\rm CDM}h^2=0.135$ \cite{WMAP}. With these numbers, 
$\delta \approx 2.1\times 10^5$. Because of this large overdensity,
UHECRs from SHDM have no GZK cutoff~\cite{BKV97}.

Clumpiness of SHDM in the halo can provide the observed small-angle
clustering~\cite{BlSh}.

The quotient $r_X=\Omega_X (t_0/\tau_X)$ of relic
abundance $\Omega_X$ and lifetime $\tau_X$ of the $X$ particle is
fixed by the observed UHECR flux as $r_X\sim 10^{-11}$. The numerical
value of $r_X$ is theoretically  calculable as
soon as a specific particle physics and cosmological model is fixed. 
In the most interesting case of gravitational production of $X$
particles, their present abundance is determined by their
mass $M_X$ and the reheating temperature $T_R$.
The life-time of the $X$ particles is on the other hand fixed by
choosing a specific particle physics model. As shown in
Ref.~\cite{tau_X}, there exist many models in which SH particles can
be quasi-stable with lifetime $\tau_X \gg 10^{10}$~yr. 
The measurement of the UHECR flux, and thereby of $r_X$, selects from
the three-dimensional parameter space $(M_X, T_R, \tau_X)$ a
two-dimensional subspace compatible with the SHDM hypothesis.
{\em Such a determination of a priori free parameters from
experimental data has nothing to do with fine-tuning}%
\footnote{In elementary-particle physics, fine-tuning is understood 
as tuning a quantity $B$ to another quantity $A$ with such precision 
that the 
predicted value $m=A-B$ is many orders of magnitude smaller than $A$
and $B$.}, a reproach sometimes used against SHDM:
Choosing a single value of $\tau_X$ from the wide range of
theoretically allowed values just reflects the present state of
theoretical ignorance.

\begin{figure}
\begin{picture}(15,9)
\put(2.5,8.3) {\epsfig{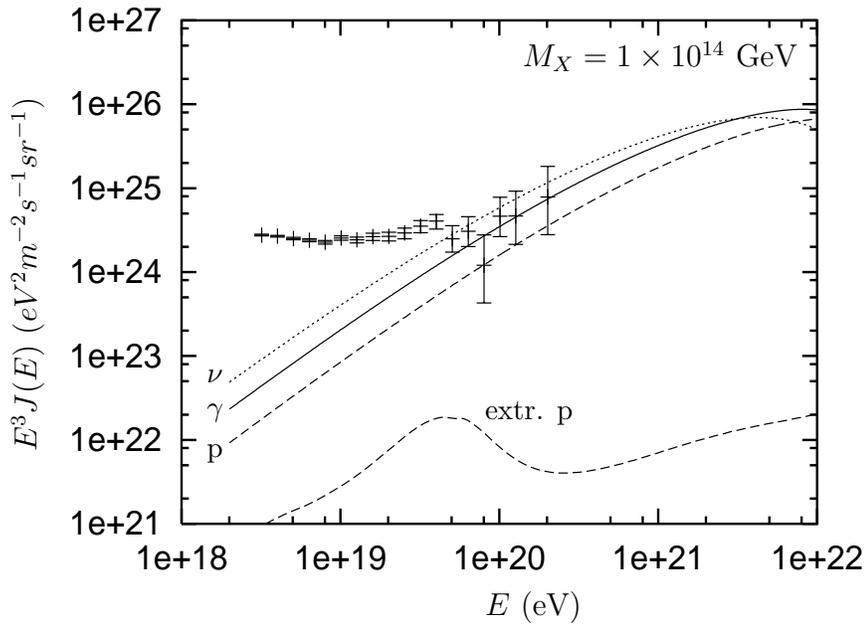} }
\put(8.0,0.2) {$E$ (eV)}
\put(2.0,4.6){
\makebox(0,0)[b]{\shortstack{$E^3 J(E)$ $(eV^2 m^{-2} s^{-1} sr^{-1})$}}%
}
\put(8.5,7.5) {$M_X=1\times 10^{14}$ GeV}
\put(4.3,3.3) {\small{\small{$\nu$}}}
\put(4.3,2.85) {\small{\small{$\gamma$}}}
\put(4.3,2.3) {\small{\small{p}}}
\put(8.0,2.8) {\small{\small{extr. p}}}
\end{picture}
\caption{\label{SHDM}
Spectra of neutrinos (upper curve), photons (middle curve) and protons
(two lower curves) in SHDM model compared with AGASA data. 
The neutrino flux is dominated by the halo
component with small admixture of extragalactic flux. The flux of
extragalactic protons is shown by the lower curve.}
\end{figure}

\begin{figure}
\begin{picture}(15,9)
\put(2.5,8.3) {\epsfig{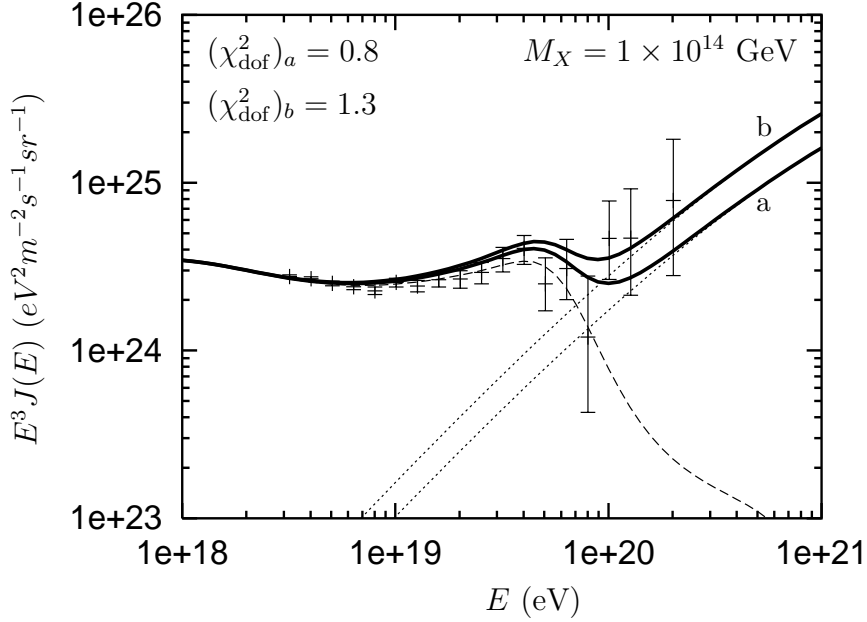} }
\put(8.0,0.2) {$E$ (eV)}
\put(2.0,4.6){
\makebox(0,0)[b]{\shortstack{$E^3 J(E)$ $(eV^2 m^{-2} s^{-1} sr^{-1})$}}%
}
\put(8.5,7.5) {$M_X=1\times 10^{14}$ GeV}
\put(4.3,7.5) {$(\chi^2_{\rm dof})_a=0.8$}
\put(4.3,6.8) {$(\chi^2_{\rm dof})_b=1.3$}
\put(11.6,6.5) {\small{b}}
\put(11.6,5.5) {\small{a}}
\end{picture}
\caption{\label{SHDM_1}
Comparison of SHDM prediction with the AGASA data. The calculated
spectrum of SHDM photons is shown by dotted curves for two different
normalizations. The dashed curve gives the spectrum of extragalactic
protons in the non-evolutionary model of Ref.~\cite{BGG03}. The sum of
these two spectra is shown by the thick curves. The $\chi^2$ values  
are given of the comparison of these 
curves with experimental data for $E\geq 4\times 10^{19}$~eV.}
\end{figure}

In Fig. \ref{SHDM}, the spectra of UHE photons, neutrinos and protons 
from the decays of SHDM particles with $M_X=1\times 10^{14}$~GeV in
the Galactic halo are presented. We have performed also a fit
to the AGASA data using the photon flux from the SHDM model and the
proton flux from uniformly disributed astrophysical sources. For the
latter we have used the non-evolutionary model of  Ref.~\cite{BGG03}. 
The photon flux is normalized to provide the best fit to the AGASA
data at $E\geq 4\times 10^{19}$~eV. The fits are shown in Fig.~\ref{SHDM_1}
with $\chi^2$/d.o.f. indicated there. 

{\em One can see  from the fits in Fig.~\ref{SHDM_1}, that the SHDM
model with the new spectra can explain only the excess of AGASA events 
at $E \gsim 1\times 10^{20}$~eV: depending on the SHDM spectrum
normalization and the details of the calculations for the extragalactic
protons, the flux from SHDM decays becomes dominant only  
above $(6-8)\times 10^{19}$~eV.} 

{\em Topological Defects\/} (for a review see~\cite{ViSh}) can naturally
produce UHE particles. The pioneering observation of this  possibility
has been made in Ref.~\cite{HiSch}.

The following TD have been discussed as potential sources of UHE
particles: superconducting strings, ordinary strings,
monopolonium (bound monopole-antimonopole pair), monopolonia
(monopole-antimonopole pairs connected by a string), networks of 
monopoles connected by strings, vortons and necklaces (see Ref.~\cite{TD}
for a review and references).

Monopolonia and vortons are
clustering in the Galactic halo and their observational signatures for
UHECR are identical to SHDM. However, as has been demonstrated in 
Ref.~\cite{BlOl}, the friction of monopolonia in cosmic plasma results in
monopolonium lifetime much shorter than the age of the universe. 

Of all other TD which are not clustering in the Galactic halo, the most 
favorable for UHECR are {\em necklaces}. Their main phenomenological 
advantage is a small separation which ensures  the arrival of highest
energy particles to our Galaxy. 

We shall calculate here the flux of UHECR from necklaces following
the works \cite{neckl,BBV98}.

Necklaces are hybrid TD produced in the symmetry breaking pattern
$G \rightarrow H \times U(1) \rightarrow H \times Z_2$. At the first
symmetry breaking monopoles are produced, at the second one each
(anti-) monopole get attached to two strings. This system resembles
ordinary cosmic strings with monopoles playing the role of
beads. Necklaces exist as the long strings and loops.

The symmetry breaking scales of the two phase transitions, $\eta_m$
and $\eta_s$, are the main parameters of the necklaces. They determine
the monopole mass, $m \sim 4\pi \eta_m/e$, and the mass of the string per
unit length $\mu \sim 2\pi \eta_s^2$. The evolution of necklaces is
governed by the ratio $r\sim m/\mu d$, where $d$ is the average
separation of a monopole and antimonopole along the string. As it is
argued in Ref.~\cite{neckl}, necklaces evolve towards configuration with
$r\gg 1$. Monopoles and 
antimonopoles trapped in the necklaces inevitably annihilate in the end,
producing heavy Higgs and gauge bosons ($X$ particles) and then
hadrons. The rate of $X$ particles production in the universe can be
estimated as \cite{neckl}
\be
\dot{n}_X \sim \frac{r^2 \mu}{t^3 M_X} \,,
\label{rate}
\ee
where $t$ is the cosmological time.

The photons and electrons from pion decays initiate e-m cascades and
the cascade energy density can be calculated as
\be
\omega_{\rm cas}=\frac{1}{2}f_{\pi}r^2\mu \int_0^{t_0} \frac{dt}{t^3}
\frac{1}{(1+z)^4}=\frac{3}{4}f_{\pi}r^2\frac{\mu}{t_0^2} \,,
\label{omega}
\ee
where $z$ is the redshift and $f_{\pi}\sim 1$ is the fraction of the 
total energy release transferred to the cascade. 

The parameters of the necklace model for UHECR are restricted by the 
EGRET observations~\cite{EGRET} of the diffuse gamma-ray flux.  
This flux is produced by UHE energy electrons and photons from
necklaces due to e-m cascades initiated in collisions with CMB photons. In the
range of the EGRET observations, $10^2 - 10^5$~MeV, the predicted spectrum
is $\propto E^{-\alpha}$ with $\alpha=2$ \cite{cascade}. The EGRET
observations determined the spectral index as $\alpha=2.10\pm 0.03$
and the energy density of radiation as $\omega_{\rm obs} \approx
4\times 10^{-6}$~eV/cm$^3$. The cascade limit consists in the 
bound $\omega_{\rm cas}\leq \omega_{\rm obs}$.

According to the recent calculations of Ref.~\cite{Strong}, the
Galactic contribution of gamma rays to the EGRET observations is
larger than estimated earlier, and the extragalactic gamma-ray
spectrum is not described by a power-law with $\alpha=2.1$. In this
case, the limit on the cascade radiation with $\alpha=2$ 
is more restrictive and is given by
\begin{equation}
\omega_{\rm cas} \leq 2\times 10^{-6} {\rm eV/cm}^3.
\label{cas}
\end{equation}
We shall use this limit in further estimates.
Using Eq.~(\ref{omega}) with  $f_{\pi}= 1$ and $t_0=13.7$~Gyr \cite{WMAP}
we obtain from Eq.~(\ref{cas}) $r^2\mu \leq 8.9\times 10^{27}$~GeV$^2$.

The important and unique feature of this TD is the small separation $D$
between necklaces. It is given by $D \sim r^{-1/2}t_0$ \cite{neckl}.
Since $r^2\mu$ is limited by e-m cascade radiation, Eq. (\ref{omega}), we
can obtain a lower limit on the separation between necklaces as
\be
D \sim \left (\frac{3f_{\pi}\mu}{4t_0^2\omega_{\rm cas}} \right )^{1/4}t_0
> 10(\mu/10^6~{\rm GeV}^2)^{1/4}~{\rm kpc} \,.
\label{separ}
\ee

This small distance is an unique property of necklaces allowing the
unabsorbed arrival of particles with the highest energies.

\begin{figure}
\begin{picture}(15,9)
\put(2.5,8.3) {\epsfig{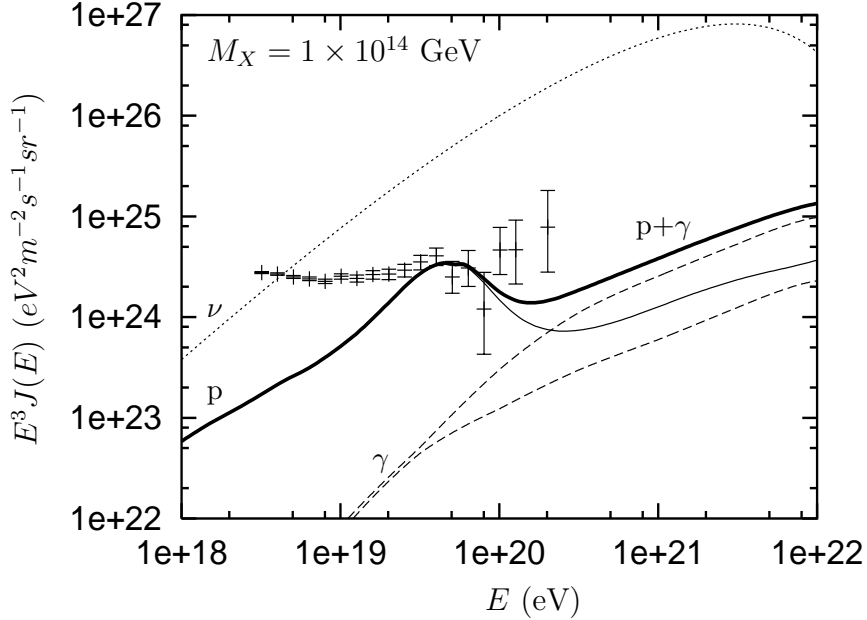} }
\put(8.0,0.2) {$E$ (eV)}
\put(2.0,4.6){
\makebox(0,0)[b]{\shortstack{$E^3 J(E)$ $(eV^2 m^{-2} s^{-1} sr^{-1})$}}%
}
\put(4.3,7.5) {$M_X=1\times 10^{14}$ GeV}
\put(4.3,4.1) {\small{\small{$\nu$}}}
\put(4.3,3.0) {\small{\small{p}}}
\put(6.5,2.1) {\small{\small{$\gamma$}}}
\put(10.0,5.2) {\small{\small{p+$\gamma$}}}
\end{picture}
\caption{\label{neck}
Diffuse spectra from necklaces. The upper curve shows neutrino flux,%
the middle - proton flux, and two lower curves - photon fluxes for
two cases of absorption. The thick continuous curve gives the sum
of the proton  and higher photon flux.}
\end{figure}

The fluxes of UHECR from necklaces are shown in Fig.\ref{neck}
(the details of these calculations will be published in a forthcoming
paper with P.~Blasi \cite{ABB03}). We used in the calculations 
$r^2\mu = 4.7\times 10^{27}$~GeV$^2$ which corresponds to 
$\omega_{\rm cas}= 1.1\times 10^{-6}$~ eV/cm$^3$, i.e. 
twice less than allowed by the bound (\ref{cas}). The mass of the $X$
particles  produced by monopole-antimonopole annihilations is taken as  
$M_X= 1\times 10^{14}$~GeV. 

From  Fig. \ref{neck} one can see that in contrast to previous 
calculations~\cite{BBV98}, the necklace model for UHECR can explain
only the highest energy part of the spectrum, with the AGASA excess somewhat
above the prediction.  This is the direct consequence of the new
spectrum of particles in $X$ decays obtained in this work. Thus UHE 
particles from necklaces can serve only as an additional component in
the observed UHECR flux. 

\section{Conclusions}
\label{conclusions}

In this paper we have compared the MC and DGLAP methods for the
calculation of hadron spectra produced by the decay (or annihilation)
of superheavy $X$ particles with masses up to $M_{\rm GUT} \sim
1\times 10^{16}$~GeV. We found an excellent agreement of these two methods.

We have further elaborated the MC simulation of Ref.~\cite{BKMC},
including the low-energy motivated hadronization functions $f_i^h(x,Q_0)$
with the properties of
fragmentation function $D_i^h(x,Q_0)$ at low scale $Q_0 \sim 1$ GeV.
Though the  new hadronization functions are somewhat different from the
old ones, all new spectra
computed for different initial partons $i$ and
different scales $M_X$ agree very well with the old spectra, as it
should be (see Section \ref{had-mc}). The small differences in the
spectra illustrate the uncertainties involved in the extraction of
hadronization functions from experimental data. However, these
uncertainties affect the spectra only at $x>0.2$. 

The calculations have been performed both for ordinary QCD and SUSY QCD. 
The inclusion of SUSY partons in the development of the cascade 
results only in small corrections, and it justifies our computation scheme 
with a single mass scale $M_{\rm SUSY}$.

In comparison to the DGLAP method, the MC simulation has the advantage of
including coherent branching. It allows reliable calculations 
at very small $x$. The Gaussian peak, the signature of the QCD spectrum,
cannot be obtained using the DGLAP equations.  

We have calculated the all-hadron spectra, as well as spectra of charged
pions and nucleons, using the DGLAP equations. Two different methods
(both based on our MC simulation) have been used.

In the first one we have used the initial fragmentation functions,
$D_q^h(x,M_Z)$ and $D_g^h(x,M_Z)$, the former determined from hadron
spectrum measured from $e^+e^-$-annihilation at $\sqrt s=M_Z$ and the
latter calculated by MC. Then we evolved these fragmentation functions
to higher scales $M_X$ with the help of the DGLAP equations.
In the second method we have used the hadronization functions 
as initial fragmentation functions,
and evolved them from the scale $Q_0 \sim 1$ GeV
to $M_X$. In the first  method the spectra calculated at different 
scales and for
different initial partons are in excellent agreement between themselves
and with MC at $x \leq 0.3$ (see Figs. \ref{MC-DGLAP1}, \ref{MC-DGLAP2}).

In the second method the agreement is equally good at scales $\sqrt s
\gg M_Z$. This also shows that hadronization functions can be seen as
fragmentation functions at low scale $Q_0 \sim 1$ GeV.
At small scales the quark singlet FF is calculated less precisely.

The disagreement at $x \sim 1$ is explained by the uncertainties in
the calculations using DGLAP and MC at large $x$, especially in the
latter. 
Using the hadronic fragmentation functions we have calculated the spectra
of photons (produced by decays of neutral pions), neutrinos (from
charged pions) and nucleons.

Our nucleon spectrum agrees well with that of Refs.~\cite{BaDr1} and
\cite{BaDr2}.

We compared also our spectrum of photons with the calculations of Ref. 
\cite{BaDr1,BaDr2,SaTo}. The comparison of the photon spectra is
interesting, because of physical reasons (photons can be observable
particles), and because the photon spectra are connected directly with the
hadron spectra. 

The spectra are in good agreement (see Section \ref{spectra} for the
detailed discussion). The disagreement at the largest $x$ is not of great 
practical interest because of the model dependent prediction of the spectrum. 
Indeed, because of the non-perturbative character of the  decay,
many-parton decays of $X$ particles can dominate over the
two parton decay considered (see Section \ref{comparison}) . Moreover,
the range $x>0.3$ corresponds to too high energies $E> 3\times 10^{12}$~GeV
at the masses of superheavy particles at interest $M_X> 1\times 10^{13}$~GeV.

We conclude that all calculations are in a good agreement especially
at small $x$ and the predicted shape of the generation spectrum
($\propto dE/E^{1.9}$)
can be considered as a signature of models with decaying (annihilating)
superheavy particles.

The predicted spectrum of SHDM model cannot fit the observed UHECR
spectrum at $1\times 10^{18}~{\rm eV}\leq E \leq (6-8)\times 10^{19}$~eV
(see Fig \ref{SHDM}). Only events at $E\gsim (6-8)\times 10^{19}$~eV, and most
notably the AGASA excess at these energies, can be explained in this model.
The robust prediction of this model is photon dominance. In present
calculations this excess diminishes to $\gamma/N \simeq 2 - 3$
(see Fig. \ref{ratio}). 

According to the recent calculation of Ref.~\cite{AhPl}, the muon content
of photon induced EAS at $E>1\times 10^{20}$ eV is high, but lower by
a factor 5 -- 10 than in hadronic showers. The muon content of EAS at
$E>1\times 10^{20}$~eV has been recently measured in
AGASA~\cite{AGASA-gamma}. The measured value
is the muon density at the distance 1000~m from the shower core,
$\rho_{\mu}(1000)$.  From 11 events at $E>1\times 10^{20}$~eV 
the muon density was measured in 6. In two of them with
energies about $1\times 10^{20}$~eV, $\rho_{\mu}$ is almost twice
higher than predicted for gamma-induced EAS. Taking into account the
contribution of extragalactic protons at this energy (see
Ref.~\cite{BGG03} for an analysis), the ratio $\gamma/p$ predicted by
the SHDM model is 1.2 -- 1.4. It is lower than  
the upper limit $\gamma/p \leq 2$ obtained by AGASA 
at $E=3\times 10^{19}$~eV on the basis of a much larger
statistics. The muon content of the remaining 4 EAS  
marginally agrees with that predicted for gamma-induced showers. The 
contribution of extragalactic protons for these events is negligible,
and the fraction of protons in the total flux can be estimated as 
$0.25 \leq p/{\rm tot}\leq 0.33$. This fraction gives a considerable
contribution to the probability of observing 4 showers with slightly
increased muon content. Not excluding the SHDM model, the AGASA events
give no evidence in favor of it.

The simultanous observation of UHECR events in fluorescent light and 
with water Cherenkov detectors has a great potential to distinguish 
between photon and proton induced EAS. An anisotropy towards the
direction of the Galactic Center is another signature of the SHDM model. 
Both kinds of informations from Auger~\cite{Auger} will be
crucial for the SHDM model and other top-down scenarios.  
 
Topological defect models are another case when short-lived superheavy
particle decays can produce UHECR. In Fig.~\ref{neck} the spectra from
necklaces are presented. One can see that at $E \gsim 1\times 10^{20}$~eV
photons dominate, and the discussion in the previous paragraph applies
here too.  In contrast to previous calculations~\cite{BBV98},
the agreement with observations is
worse: necklaces can explain only the highest energy part of the spectrum 
in Fig.~\ref{neck}, with the AGASA excess somewhat above the prediction.  
In the other energy ranges, 
UHE particles from necklaces can provide only a subdominant
component. Other TDs suffer even more problems (see Ref.~\cite{BBV98}).

\section*{Acknowledgments}
We thank Pasquale Blasi for participation in 
the calculation of the spectra from necklaces. Cyrille
Barbot  and Ramon Toldr\`a are thanked for sending us their results
and for helpful comments. Finally, we gratefully
acknowledge useful discussions with Giuseppe Di Carlo, Motohiko Nagano, 
Sergey Ostapchenko, and in particular with Valery Khoze. 
MK is grateful to the Deutsche
Forschungsgemeinschaft (DFG) for an Emmy Noether fellowship. 
This work has been performed within the INTAS project 99-01065.

\newpage

\begin{table}
\begin{center}
\caption{Splitting functions for QCD and SUSY QCD from \cite{SUSY} 
($C_F=4/3$ and $N_C=3$)}
\label{sf}
\vskip 0.5cm
\begin{tabular}{c|c}
\hline
quark & gluon \\
\hline 
   &  \\
$P^{(0)}_{qq}(x)=C_F\left [\frac{2}{1-x}-1-x \right]$ &
$P^{(0)}_{qg}(x)=\frac{1}{2}\:\left [x^2+(1-x)^2 \right]$ \\
   &  \\
$P^{(0)}_{gq}(x)=C_F\left [\frac{1+(1-x)^2}{x}\right]$ &
$P^{(0)}_{gg}(x)=2N_C\left [\frac{1}{1-x}+\frac{1}{x}+x(1-x)-2 \right]$ \\
   &  \\
$P^{(0)}_{\tilde{q}q}(x)=C_F x$ &
$P^{(0)}_{\tilde{q}g}(x)=\frac{1}{2}\:\left [2x(1-x)\right ]$ \\
   &  \\
$P^{(0)}_{\tilde{g}q}(x)=C_F \left [1-x\right ]$ &
$P^{(0)}_{\tilde{g}g}(x)=2N_C\left [x^2+(1-x)^2\right ]$ \\
   &  \\
\hline
squark & gluino \\

\hline
   &  \\
$P^{(0)}_{q\tilde{q}}(x)=C_F$ &
$P^{(0)}_{q\tilde{g}}(x)=\frac{1}{2}\:\left [1-x \right]$ \\
   &  \\
$P^{(0)}_{g\tilde{q}}(x)=C_F\left [2\frac{1-x}{x}\right]$ &
$P^{(0)}_{g\tilde{g}}(x)=N_C\left [\frac{1+(1-x)^2}{x}\right]$ \\
   &  \\
$P^{(0)}_{\tilde{q}\tilde{q}}(x)=C_F \left [2\frac{x}{1-x} \right ]$ &
$P^{(0)}_{\tilde{q}\tilde{g}}(x)=\frac{1}{2}\: x$ \\
   &  \\
$P^{(0)}_{\tilde{g}\tilde{q}}(x)=C_F$ &
$P^{(0)}_{\tilde{g}\tilde{g}}(x)=N_C\left [\frac{1+x^2}{1-x}\right ]$ \\
   &  \\
\end{tabular}
\end{center}
\end{table}

\unitlength1.0cm

\newpage
\phantom{}

\unitlength1.0cm

\unitlength1.0cm

\unitlength1.0cm

\unitlength1.0cm

\unitlength1.0cm

\unitlength1.0cm

\unitlength1.0cm

\unitlength1.0cm

\unitlength1.0cm

\unitlength1.0cm

\unitlength1.0cm

\phantom{}
\newpage

\unitlength1.0cm

\unitlength1.0cm

\unitlength1.0cm

\unitlength1.0cm

\unitlength1.0cm

\unitlength1.0cm

\unitlength1.0cm

\unitlength1.0cm

\unitlength1.0cm

\end{document}